\documentclass[11pt]{article}
\pdfoutput=1 

\usepackage{jheppub}
                     
\usepackage{amsmath}
\usepackage{amssymb}
\usepackage{dsfont}
\usepackage{bm}
\usepackage[T1]{fontenc}
\usepackage{caption}
\usepackage{subcaption}
\usepackage{color}
\usepackage[export]{adjustbox}
\usepackage[normalem]{ulem}
\usepackage{xcolor}
\usepackage{tocvsec2}

\def\nn{\nonumber}
\def\({\left(}
\def\){\right)}
\def\[{\left[}
\def\]{\right]}

\newcommand{\beq}{\begin{equation}}
\newcommand{\beqn}{\begin{eqnarray}}
\newcommand{\eeq}{\end{equation}}
\newcommand{\eeqn}{\end{eqnarray}}
  \newcommand{\be}{\begin{equation}}
  \newcommand{\ee}{\end{equation}}
  \newcommand{\bea}{\begin{eqnarray}}
\newcommand{\eea}{\end{eqnarray}}
\newcommand{\Beq}{\begin{equation}\begin{aligned}}
\newcommand{\Eeq}{\end{aligned}\end{equation}}

\newcommand{\la}{\langle}
\newcommand{\ra}{\rangle}
\newcommand{\bra}[1]{\langle{#1}|}
\newcommand{\ket}[1]{|{#1}\rangle}
\renewcommand{\it}{\textit}
\newcommand{\bt}{\textbf}

 \newcommand{\axion}{\phi}

\definecolor{darkgreen}{cmyk}{0.85,0.2,1.00,0.2}
\newcommand{\peter}[1]{\textcolor{red}{[{\bf PA}: #1]}}

\newcommand{\pushright}[1]{\ifmeasuring@#1\else\omit\hfill$\displaystyle#1$\fi\ignorespaces}
\newcommand{\pushleft}[1]{\ifmeasuring@#1\else\omit$\displaystyle#1$\hfill\fi\ignorespaces}

\definecolor{darkgreen}{cmyk}{0.85,0.2,1.00,0.2}

\title{\boldmath Production and backreaction of massive fermions during axion inflation with non-Abelian gauge fields}

\makeatother

\author[a]{Peter Adshead,}
\affiliation[a]{Illinois Center for Advanced Study of the Universe and Department of Physics, University of Illinois at Urbana-Champaign, Urbana, IL 61801, USA }
\author[b]{Aike Liu,}
\affiliation[b]{Walter Burke Institute for Theoretical Physics, California Institute of Technology, Pasadena, CA 91125}
\author[a]{and Kaloian D. Lozanov}

\emailAdd{adshead@illinois.edu}
\emailAdd{aliu7@caltech.edu}
\emailAdd{klozanov@illinois.edu}

\abstract{We study the production and backreaction of massive vector-like fermions in the background of a classical SU(2) gauge field during axion-driven inflation. We demonstrate all ultraviolet divergences due to the interactions with the fermions can be absorbed by renormalization of the axion wavefunction and the gauge coupling. The effects of the fermion-axion interaction vanish in the massless limit as required by symmetry. For very massive fermions, contact interactions are induced between the axion, the gauge field and the gravitational field. In this massive limit, we find the usual axion-gauge field interactions are induced, however, in addition we observe the appearance of axion self-interactions, as well as  kinetic braiding of the axion with the Einstein tensor. These new axion derivative interactions present intriguing opportunities for model building and phenomenology.}

\begin{document}
\maketitle
\flushbottom

\section{Introduction}

The inflationary paradigm \cite{Guth:1980zm, Linde:1981mu, Albrecht:1982wi} is both a phenomenological success and an enduring theoretical challenge. Observations of a red-tilted spectrum of Gaussian, adiabatic density fluctuations in the cosmic microwave background (CMB) are consistent with a period of early accelerated expansion (see, e.g., \cite{Ade:2015lrj}). Furthermore, observations of polarization are becoming increasingly accurate, and are beginning to rule out regions of parameter space favored by large-field inflation, and currently limit the tensor-to-scalar ratio, $r \lesssim 0.04$ \cite{BICEP:2021xfz}. Upcoming experiments will further reduce the limit, or discover primordial gravitational waves \cite{CMB-S4:2016ple, Shandera:2019ufi, Abazajian:2019eic}. 

In standard inflationary scenarios, the tensor-to-scalar ratio is a direct probe of the energy scale of inflation. The amplitude of the primordial gravitational wave background, and thus the primordial B-mode polarization, is directly proportional to the Hubble rate during inflation, and is therefore a  probe of the inflationary energy scale. In particular, observable gravitational waves (via B-modes in the CMB) require inflation to happen at, or near to the energy scale associated with grand unification of the gauge couplings, the so-called "GUT-scale" $\sim 10^{16}$ GeV. Further, these models also imply that the field driving the inflationary expansion must move over a distance that is large compared to the Planck scale \cite{Lyth:1996im}.

However, in models involving gauge fields, observable (in CMB B-mode polarization) gravitational waves can be produced at significantly lower energy scales, while the distances the fields move over are small compared to the Planck scale \cite{Adshead:2013qp, Dimastrogiovanni:2012ew}. In these scenerios,  non-Abelian gauge fields in classical configurations play a key role in the evolution either in driving the accelerated expansion directly---Gaugeflation \cite{Maleknejad:2011jw, Maleknejad:2011sq}---or by coupling to an axion-inflaton and facilitating slow-roll on a steep potential---Chromo-Natural inflation \cite{Adshead:2012kp, Adshead:2012qe, Martinec:2012bv}.\footnote{Scenarios with Abelian gauge fields have been considered previously, for example, \cite{Sorbo:2011rz, Namba:2015gja}. However, at least when the gauge fields are coupled to the inflaton,  the strongest gravitational wave production occurs during the latter stages of inflation and during reheating. In particular, gravitational wave production during gauge-preheating is so prolific that these models are already constrained by measurements of the effective number of neutrino species in the CMB, $N_{\rm eff}$ \cite{Adshead:2018doq,Adshead:2019lbr, Adshead:2019igv}.}

While Chromo-Natural inflation and Gauge-flation are in tension with current data \cite{Dimastrogiovanni:2012ew, Adshead:2013qp, Adshead:2013nka, Namba:2013kia, Fujita:2022fff}, they can be brought into agreement by flattening the potential \cite{Maleknejad:2016qjz, Caldwell:2017chz, Papageorgiou:2018rfx}, or by giving the gauge fields masses \cite{Adshead:2016omu, Adshead:2017hnc}. More broadly, a spectator Chromo-Natural sector (involving a spectator axion field coupled to classical SU(2) gauge fields), or Gauge-flation sector \cite{Iarygina:2021bxq} can generate primordial gravitational waves at energy scales far below the scale of grand unification \cite{Dimastrogiovanni:2016fuu}. Detailed analysis has revealed that these spectator models are not without problems. The resulting gravitational waves are highly non-Gaussian \cite{Agrawal:2017awz}, and the parameter space is already constrained by data \cite{Agrawal:2018mrg, Thorne:2017jft}. Further, the backreaction of the fluctuations onto the curvature perturbation also places restrictions on the allowed space of models \cite{Papageorgiou:2019ecb}. However, given the importance of primordial gravitational waves as so-called smoking gun signatures of inflation, it is important to investigate in detail any scenarios that potentially produce gravitational waves that run counter to the standard lore.

The classical non-Abelian gauge field configuration utilized in models such as Chromo-Natural inflation leads to remarkable phenomenology. In particular, the embedding of the gauge field into the background spontaneously breaks parity, which leads to striking parity-violating phenomenology of quantum fields propagating on the classical field background. In this work we study the phenomenology of fermions propagating on the classical gauge field background. Previous studies in this area have focussed on the massless limit \cite{Domcke:2018gfr}, as well as on the production and backreaction of massive fermions \cite{Mirzagholi:2019jeb, Maleknejad:2019hdr, Maleknejad:2020yys, Maleknejad:2020pec}. The production, backreaction and phenomenology of massive fermions  during axion inflation has also been recently studied in Refs.\ \cite{Adshead:2015kza, Adshead:2015jza, Adshead:2018oaa, Adshead:2019aac, Wang:2019gbi,  Roberts:2021plm}. Finally, the axion-assisted Schwinger effect was recently studied in Refs.\ \cite{Domcke:2021fee, Kitamoto:2021wzl}.

In this work we revisit massive fermion production, paying careful attention to the quantization and symmetry of the theory under charge conjugation. We additionally provide new analytic solutions in the massless limit, and include the effect of couplings to the axion. We regularize the axial and gauge backreaction currents using Pauli-Villars regulator fields, and demonstrate that all remaining ultraviolet (UV) divergences can be absorbed by renormalizing the gauge coupling and the axion wavefunction. In the massless limit, the axion backreaction vanishes, as required by chiral symmetry. In this limit, the fermions renormalize the gauge coupling and lead to small corrections to the gauge field equation of motion. In the opposite limit, where the fermions are very massive, the physical backreaction currents vanish leaving effective contact interactions between the axion and gauge fields,  the axion and the spacetime curvature, and axion self-interactions.

This paper is organized as follows. In section \ref{sec:singquant} we introduce the model, together with the background and our conventions.  In section \ref{sec:eomandquant} we quantize the fermions in the classical SU(2) gauge field background, and solve their classical equations of motion. In section  \ref{sec:currentsandbackreact} we compute the axial and gauge currents and the associated backreaction on the axion and gauge field equations of motion. Finally, we conclude in section \ref{sec:conclusions}. Many details of our lengthy and technical computations are relegated to appendices. In appendix \ref{app:antisym}, we demonstrate that antisymmetrization of quadratic fermion-field operators is not necessary when computing expectation values. Appendix \ref{app:WKB} details our method for finding the adiabatic solutions to the field equations while  appendix \ref{app:seriessols} contains details of our series solutions to the equations of motion for the coupled fermion states, and appendix \ref{app:PVsols} details the method of finding adiabatic solutions in the large mass limit. Finally, in appendix \ref{app:analints}, we provide details of the analytic integrations of the fermion currents.

We work in units where $\hbar = c = k_B = 1$, and denote by $M_{\rm Pl} = 2.435\times  10^{18}$ GeV the reduced Planck mass.
%
\section{The model, background, and conventions}\label{sec:singquant}
%
We consider a theory containing a slowly-rolling axion field coupled to a non-Abelian gauge field, minimally coupled to Einstein gravity\footnote{We work in mostly-minus metric convention, and use Greek letters to denote 4-dimensional spacetime indices. Roman letters from the start of the alphabet are used to denote SU(2) gauge indices, while Roman letters from the middle of the alphabet denote spatial spacetime indices. Further, capital Roman letters from the start of the alphabet denote 4D Lorentz indices and $\eta^{AB} = {\rm diag}[1, -1, -1, -1]$ is the flat Minkowski metric.}
\be\label{eqn:backgroundaction}
S = \int d^4 x \sqrt{-g} \[\frac{M_{\rm Pl}^2}{2} R -\frac14 F_{\mu\nu}^a  F^{a\mu\nu} +\frac{g^{\mu \nu}}{2} \partial_\mu \axion
 \partial_\nu \axion - V(\axion) + \frac{\alpha \axion}{4 \pi f} F_{\mu\nu}^a  \tilde F^{a\mu\nu}\],
 \ee
 where $\alpha = g^2/(4\pi)$, $g$ is the gauge coupling, and $f$ is a mass-scale associated with the axion.

 We have in mind scenarios such as Chromo-Natural inflation \cite{Adshead:2012kp}, however, our analysis equally applies to scenarios where the axion-gauge field sector is merely a spectator and is not responsible for generating the background quasi-de Sitter spacetime, such as that first proposed in Ref.\ \cite{Dimastrogiovanni:2016fuu}. We take the gauge field to be in the classical, flavor-space locked configuration by identifying the gauge indices with spatial indices of the spatial tetrad $e^a_{~i}$
\be
A^a_0 = 0, \quad A^a_i = Q e^a_{~i}, \quad g_{\mu\nu} = \eta_{AB}e^A_{~\mu}e^B_{~\nu}.
\ee
We assume a background Friedman-Robertson-Walker metric, so that $e^a_{~i} = a \delta^{a}_{~i}$ on the background spacetime, where $a$ is the scale factor. We assume that both the gauge field and the axion to be in homogeneous and slowly evolving configurations satisfying
\begin{align}
\delta_Q \equiv \frac{ \dot{Q}}{HQ} \ll1 ,\quad \epsilon_{\dot\axion}  \equiv \frac{\dot{\axion}^2}{2 H^2 M_{\rm Pl}^2} \ll 1.
\end{align}
We further assume that $ \dot{\axion}\approx {\rm const.}$, and define 
\begin{align}
\xi = \frac{\lambda \dot{\axion}}{2fH}, \quad m_{Q} = \frac{g Q}{H},
\end{align}
which we take to be free parameters. Here and throughout an overdot denotes a cosmic time derivative, and $H = \dot{a}/a$ is the Hubble rate.

In addition to the classical field content we couple the theory to a vectorlike Dirac fermion doublet charged under the $SU(2)$ gauge field  
\be\label{eq:fermact1}
 S_{\rm f} = \int  dx^4 \sqrt{-g}\left[i \overline Y(  \gamma^\mu D_\mu ) Y - m  \overline Y e^{-i \frac{2 \lambda \phi}{f}\gamma_5} Y \right].
 \ee
We have also introduced a coupling of the fermion to the axion that might arise from the spontaneous breaking of the axial symmetry in the UV, and we have allowed the axion decay constant to differ by introducing a factor of $\lambda$ which parameterizes the ratio.
 
The gauge-covariant derivative is 
\be
\gamma^\mu D_\mu = \gamma^C e_C^{~\mu}\(  \partial_\mu + w_{AB\mu} \Sigma^{AB}  - ig A_\mu^a \tau^a\),
\ee
where $\tau^a =  \sigma^a/2$ are the generators of SU(2), and $\sigma^a$ are the Pauli matrices.  The spin connection $w^{A}_{~B\mu}$ is given by
\be
w^{A}_{~B\mu} = e^A_{~\nu} \nabla_\mu (e^{\nu}_{~ B})= e^A_{~\nu} \partial_\mu (e^{\nu}_{~ B}) +  e^A_{~\nu}  e^\sigma_{~B} \Gamma^{\nu}_{~\sigma \mu} ,
\ee
where $ \Gamma^{\nu}_{~\sigma \mu} $ are the usual Christoffel symbols,  and  
\be
\Sigma^{AB}= \frac14 \[\gamma^A, \gamma^B\],
\ee
are the generators of the Lorentz group. We work with the Clifford algebra 
\begin{align} 
\{ \gamma^A,  \gamma^B\} = 2\eta^{AB},
\end{align} 
and the Weyl basis for the gamma matrices
\begin{align}
\gamma^0 = & \(\begin{matrix}0 & \mathds{1}\\ \mathds{1} & 0 \end{matrix}\), \quad \gamma^i = \(\begin{matrix}0 & \sigma^i\\ -\sigma^i & 0 \end{matrix}\), \quad \gamma^5 = \(\begin{matrix} -\mathds{1} & 0 \\ 0 & \mathds{1}\end{matrix}\).
\end{align}

We are primarily interested in the production of fermions and their backreaction on the homogeneous mode, and in what follows we ignore fluctuations of the metric. After rescaling and rotating the fermions by 
\begin{align}\label{eqn:fermrot}
\psi = a^{3/2} e^{-i\lambda\frac{\phi}{f}\gamma_5}Y,
\end{align}
and taking the axion background to be homogeneous, the fermion action in eq.\ \eqref{eq:fermact1} reads\footnote{Note that the fermion doublet, $\psi$, has eight components. Since the $SU(2)$ structure is rather trivial (other than the gauge interaction)  rather than cluttering the notation by explicitly writing expressions such as $\mathds{1}_{2\times 2}\otimes \gamma^\mu$, we suppress the direct products except where necessary.}
  \begin{align}\label{eqn:fermionaction}
 S_{\rm f}
&  = \int  d^3x d\eta  \left[ i \bar \psi  \gamma^\mu \( \partial_\mu -i gA^a_\mu \tau^a \) \psi - a m  \bar \psi \psi - \frac{\lambda \partial_\eta \phi}{f} \bar \psi \gamma^0 \gamma^5 \psi\right ],
 \end{align}
where $\eta$ is conformal time. 

As demonstrated in Ref.\ \cite{Adshead:2021ezw}, the path integral measure is not invariant under the field redefinition in eq.\ \eqref{eqn:fermrot}. This redefinition leads to the additional contributions to the lagrangian
\begin{align}\label{eqn:lagrangianshift}\nn
\Delta \mathcal{L} = &  \hbar\sqrt{-g}\Big[\frac{\lambda^2}{12\pi^2}\frac{G^{\mu\nu}\partial_\mu\phi\partial_\nu\phi}{f^2} - \frac{\lambda^2}{12\pi^2}\frac{(\Box\phi)^2}{f^2} - \frac{\lambda^4}{6 \pi^2}\(\frac{\partial_\mu \phi \partial^\mu\phi}{f^2}\)^2 - \frac{\alpha}{2\pi}\frac{\lambda \phi}{f} F_{\mu\nu}^a  \tilde F^{a\mu\nu} \\ & \qquad \qquad\qquad+\frac{1}{192\pi^2}\frac{\lambda \phi}{f} \epsilon^{\mu\nu\alpha\beta}R^{\rho\sigma}{}_{\mu\nu}R_{\rho\sigma\alpha\beta}\Big],
\end{align}
where these arise at one-loop, and we have explicitly restored the factor of $\hbar$ to indicate this. We demonstrate below that the addition of these terms is critical to recovering the correct behavior of the theory in the massless limit  $m \to 0$ where the axion-fermion interaction in the action in eq.\ \eqref{eq:fermact1} disappears.

\section{Fermion equations of motion and quantization}\label{sec:eomandquant}

In this section we discuss the properties of the equations of motion of the fermions under spatial rotations. We expand the field into modes, quantize, and make use of charge-conjugation to simplify the solutions. We then solve the classical equations of motion.

\subsection{Rotations and gauge transformations}

Variation of the action in eq.\ \eqref{eqn:fermionaction} with respect to $\psi$ yields the equations of motion for the fermion doublet
\begin{align}
\[ i  \gamma^\mu  \partial_\mu+ g \gamma^\mu A^a_\mu \tau^a- a m - \frac{\lambda \partial_\eta \phi}{f}  \gamma^0 \gamma^5\] \psi(\eta, {\bf x} ) = 0.
\end{align}
After Fourier transforming 
\begin{align}
\psi(\eta, {\bf x} ) = \int \frac{d^{3}k}{(2\pi)^{3/2}}\psi_{\bf k}(\eta)e^{i{\bf k}\cdot{\bf x}},
\end{align}
the equations of motion read
\begin{align}\label{eq:diracfourier}
\[ i  \gamma^0 \partial_0 - \gamma^i k_{i} + g \gamma^\mu A^a_\mu \tau^a- a m - \frac{\lambda \partial_\eta \phi}{f}  \gamma^0 \gamma^5\]\psi_{\bf k}(\eta)= 0.
\end{align}
Notice that this equation is not rotationally invariant due to the presence of the gauge field. This can be seen explicitly as follows. Consider a mode, ${\bf p}$, that is related to the mode ${\bf k}$ by a rotation about an axis $\hat{n}$ by an angle $\theta$ , 
\begin{align}\label{eq:rotationdef}
p_i = \mathcal{R}^j_{i}(\theta, \hat{n})k_j= \exp\[-\theta\epsilon_{ijk}\hat{n}^k\] k_j,
\end{align}
where  $\epsilon_{ijk}$ is the Levi-Cevita symbol.  The mode with momentum ${\bf p}$ is a solution of the equation
\begin{align}\label{eq:diracrot}
\[ i  \gamma^0 \partial_0 - \gamma^i  \mathcal{R}^j_{i}k_j + g \gamma^\mu A^a_\mu \tau^a- a m - \frac{\lambda \partial_\eta \phi}{f}  \gamma^0 \gamma^5\]\psi_{  {\bf p}}(\eta)= 0.
\end{align}
Denoting the spinor representation of the Lorentz transformation (parameterized by $\omega_{AB}$) by 
\begin{align}
\Lambda_{\frac{1}{2}}= \exp\[\frac{1}{2}\omega_{AB}\Sigma^{AB}\],
\end{align}
we use the relation (see, e.g., \cite{Peskin:1995ev}), 
\begin{align}
\Lambda^{-1}_{\frac{1}{2}}(\theta, \hat{n})\gamma^{i}\Lambda_{\frac{1}{2}}(\theta, \hat{n}) = \mathcal{R}^{i}{}_{j}(\theta, \hat{n})\gamma^{j}, \quad \text{ where }\quad   \omega_{ij} = \theta\epsilon_{ijk}\hat{n}^k,
\end{align}
and $\mathcal{R}$ is the rotation in eq.\ \eqref{eq:rotationdef}. Multiplying eq.\ \eqref{eq:diracrot} by $\Lambda^{-1}_{\frac{1}{2}}(-\theta, \hat{n})$, we obtain
\begin{align}\label{eq:diracrot2}
\[ i  \gamma^0 \partial_0 - \gamma^i  k_i + g \gamma^0 A^a_0 \tau^a+ g \gamma^i  \mathcal{R}^{j}{}_{i}(-\theta, \hat{n}) A^a_j \tau^a- a m - \frac{\lambda \partial_\eta \phi}{f}  \gamma^0 \gamma^5\]\Lambda^{-1}_{\frac{1}{2}}(-\theta, \hat{n})\psi_{ {\bf p}}(\eta)= 0.
\end{align}
This expression implies that the solutions of the Dirac equation in this case depend non-trivially on direction.  However, the fields $\psi$ are not gauge invariant. The physical  quantities of interest are gauge invariant, and thus we are  free to make ($SU(2)$) gauge transformations on the above equations. Under a (global) $SU(2)$ rotation the gauge potential transforms as
\begin{align}
\bt{A}_{\mu} \to \tilde{\bt{A}}_{\mu} = U(\alpha) \bt{A}_{\mu} U^{\dagger}(\alpha ) = \mathcal{U}_{ab} A^a_{\mu}\tau^b, \quad U = \exp\(i\alpha^a \tau^a\), \quad \mathcal{U}_{ab} = \exp\(-\epsilon_{abc}\alpha^c\).
\end{align}
Notice that since $A^a_{i}  \propto \delta^{a}_i$,  if we choose the parameters $\alpha^a = -\theta \hat{n}^i \delta^a_i$, where $\theta$ and $\hat{n}$ are the axis and angle of rotation that takes ${\bf k} \to {\bf p}$, the gauge transformation undoes the effect of the rotation in eq.\ \eqref{eq:diracrot2}. Therefore,  $U(-\theta \hat{n}^i \delta^a_i) \Lambda^{-1}_{\frac{1}{2}}(-\theta, \hat{n})\psi_{\bf p}(\eta)$ solves the same equation of motion as $\psi_{\bf k}(\eta)$.  

All  quantities of interest are invariant under simultaneous rotations and gauge transformations. This implies that we may simply analyze a single momentum, which we take to be the $\hat{z}$ direction. The field for any general momentum ${\bf k}$ is then found by a simultaneous spatial and gauge rotation.
For a  general direction $\hat{ k} =  \hat{k}(\phi, \theta, \gamma)$, where $\phi$, $\theta$, and $\gamma$ are the Euler angles defined by
\begin{align}
\hat{k}^i =  R^{i}_{j}(\phi, \hat{z})R^{j}_{l}(\theta, \hat{y})R^{l}_{m}(\gamma, \hat{z})\hat{z}^m,
\end{align}
the transformation is explictely given by
\begin{align}\label{eq:rotation}
\psi_{\bf k}(\eta) =\mathcal{G}_{\hat{z} \to \hat{k}}  \psi_{ k\hat{z}}(\eta),
\end{align}
where
\begin{align}
\mathcal{G}_{\hat{z} \to \hat{k}} = U(\hat{z},\phi)U(\hat{y},\theta) U(\hat{z},\gamma)\Lambda_{\frac{1}{2}}(\hat{z},-\gamma)\Lambda_{\frac{1}{2}}(\hat{y},-\theta)\Lambda_{\frac{1}{2}}(\hat{z},-\phi).
\end{align}

\subsection{Charge conjugation and quantization}

To quantize the theory, we expand the field into creation and annihilation operators
\begin{align}\label{eq:modeexpansion}
\psi(\eta, {\bf x})  = \sum_{i=1}^4\int \frac{d^3k}{(2\pi)^{3/2}}\[ e^{i \bt{k} \cdot \bt{x} }  U_{i}(\eta,\bt{k} ) a_i (\bt{k})+ e^{-i \bt{k} \cdot \bt{x} }  V_i (\eta,\bt{k} )b^\dag_i (\bt{k})\],
\end{align}
where the sum runs over the independent solutions of the equation of motion. The creation and annihilation operators satisfy the anti-commutation relations
\begin{align}
\{a_i (\bt{k}), a^\dagger_j (\bt{k}')\} = \delta_{ij}\delta^{3}({\bf k} - {\bf k'}), \quad \{b_i (\bt{k}), b^\dagger_j (\bt{k}')\} = \delta_{ij}\delta^{3}({\bf k} - {\bf k'}), \\
\{a_i (\bt{k}), a_j (\bt{k}')\} =  \{b_i (\bt{k}), b_j (\bt{k}')\} = \{a_i (\bt{k}), b^\dagger_j (\bt{k}')\} = \{b_i (\bt{k}), a^\dagger_j (\bt{k}')\} =  0.
\end{align}
Imposing the canonical anti-commutation relation between $\psi$ and its canonical momenta $i\psi^\dagger$,
\begin{align}
\{\psi_{\alpha}(\eta, {\bf x}), \psi^\dagger_{\beta}(\eta, {\bf y})\} = \delta_{\alpha\beta}\delta^{3}({\bf x}-{\bf y})
\end{align}
leads to the condition
\begin{align}\label{eq:quantnorm}
\sum_{i = 1}^4\[U_{i}(\eta,\bt{k} ) U^\dagger_{i}(\eta,\bt{k} )+V_{i}(\eta,-\bt{k} ) V^\dagger_{i}(\eta,-\bt{k} ) \]_{\alpha\beta} = \delta_{\alpha\beta}.
\end{align}

The fields $U_{i}(\eta,\bt{k} )$ and $V_{i}(\eta,\bt{k} )$ are not independent, but rather are related to each other via charge conjugation. On the one hand, under charge conjugation, the creation and annihilation operators transform as
\begin{align}\label{eqn:hilbertcc}
{\bf C}a_i (\bt{k}){\bf C}^{-1} = b_i (\bt{k}), \quad {\bf C}b_i (\bt{k}){\bf C}^{-1} = a_i (\bt{k}).
\end{align}
On the other hand, the charge conjugation operator is an anti-linear transformation of the fields which takes
\begin{align}\label{eqn:su2cc}
\psi \to \psi^c = -2i\tau^2\otimes \tilde{C}(\bar{\psi})^T,
\end{align}
where $\tilde{C}  = i\gamma^2\gamma^0$ is the usual Dirac charge-conjugation operator, and $\tau^2 = \sigma^2/2$ is the $SU(2)$ generator.\footnote{ Note that the form of charge conjugation we use here is different than that used in the absence of the background SU(2) gauge field. For a non-Abelian gauge theory coupled to a fermion current, charge conjugation is the discrete symmetry \cite{Tyutin:1982fx}
\begin{align}
\psi \to C \psi^{*}, \quad A_{\mu} \to A_{\mu}^{T}
\end{align}
which leaves the combination $\bar{\psi}A_{\mu}\gamma^\mu\psi$ invariant. In the case at hand, because the background gauge field is fixed, we instead perform the additional active rotation on the fermion fields $\psi \to i\sigma_2 C \psi^*$.} 
It is straightforward to check that $\psi^c$ solves the same equation of motion as $\psi$. 

Taken together, eqs.\ \eqref{eqn:hilbertcc} and \eqref{eqn:su2cc} imply that 
\begin{align}
V_{i}(\eta,\bt{k} ) =  -2i\tau^2\otimes \tilde{C} \gamma^0 U^*_{i}(\eta, {\bf k}).
\end{align}
In order to impose the condition in eq.\ \eqref{eq:quantnorm}, it is convenient to rewrite eq.\ \eqref{eq:modeexpansion} as 
\begin{align}\label{eq:modeexpansion2}
\psi(\eta, {\bf x})  = \sum_{i=1}^4\int \frac{{\rm d}^3k}{(2\pi)^{3/2}}e^{i \bt{k} \cdot \bt{x} }  \[  U_{i}(\eta,\bt{k} ) a_i (\bt{k})+  V_i (\eta, - \bt{k} )b^\dag_i (-\bt{k})\].
\end{align}
We choose our reference momenta along the $\hat{z}$ direction, whereby
\begin{align}\label{eq:modeexpansion3}
\psi(\eta, {\bf x})  = \sum_{i=1}^4\int \frac{d^3k}{(2\pi)^{3/2}}e^{i \bt{k} \cdot \bt{x} } \mathcal{G}_{\hat{z}\to \hat{k}} \[  U_{i}(\eta,k \hat{z} ) a_i (\bt{k})+ \tilde{V}_{i}(\eta, { k}\hat{z}) b^\dag_i (-\bt{k})\],
\end{align}
where 
\begin{align}\label{eq:rotatedantispinor}
\tilde{V}_{i}(\eta, { k}\hat{z}) = \mathds{1}_{2\times2}\otimes \mathcal{G}_{\hat{z} \to - \hat{z}}(-i\sigma^2)\otimes( \tilde{C} \gamma^0) U^*_{i}(\eta, { k}\hat{z}) = -\mathds{1}_{2\times2}\otimes\gamma^5\gamma^0U^*_{i}(\eta, { k}\hat{z}) .
 \end{align}
 
\subsection{Mode equations and classical solutions}\label{sec:classsols}

We further expand the modes into helicity states
\begin{align}\label{eq:modeeqns}
U_{ai}(\eta,k \hat{z} ) = \frac{1}{\sqrt{2}}\sum_{r = \pm}\(\begin{matrix} u_{ai}^{r}(k, \eta)\chi_{r}(\hat{z}) \\ rv_{ai}(k, \eta) \chi_r(\hat{z}) \end{matrix}\), \quad \tilde{V}_{ai}(\eta, { k}\hat{z})  = \frac{1}{\sqrt{2}}\sum_{r = \pm} \(\begin{matrix}  rv^{*r}_{ai}(k, \eta) \chi_r(\hat{z})\\ -u^{*r}_{ai}(k, \eta)\chi_{r}(\hat{z})  \end{matrix}\),
\end{align}
where $ a \in \{1, 2\}$ are the $SU(2)$ indices of the fermion doublet, and $\chi_{r}(\hat{z})$ are helicity spinors. As above, the subscript $i \in \{1, 2, 3, 4\}$ denotes the independent solutions to the classical equations of motion. Since our momentum is oriented in the $\hat{z}$ direction, the helicity spinors are simply given by
\begin{align}
\chi_{+}(\hat{z}) \equiv \chi_+ = \(\begin{matrix} 1\\0\end{matrix}\), \quad \chi_{-}(\hat{z}) \equiv \chi_-= \(\begin{matrix}0\\ 1 \end{matrix}\).
\end{align}
Substituting eq.\ \eqref{eq:modeeqns} into the Fourier space Dirac equation, eq.\ \eqref{eq:diracfourier}, leads to the equations for each helicity $r = \pm$
\begin{align}\nn
& \(\begin{matrix} - am & i \partial_0 - r\({k}+ 2r aH \xi- \frac{gaQ}{2}\)  \\  i \partial_0 +r \(k +2r aH \xi- \frac{gaQ}{2}\)  & -am\end{matrix} \)\(\begin{matrix} u^r_1 \\ rv^r_1 \end{matrix}\) \\ &  \quad\qquad \quad\qquad\quad\qquad \quad\qquad \quad\qquad \quad\qquad =  -\frac{gaQ}{2}\(\begin{matrix} 0 & (1- r) \\ -(1- r)  & 0 \end{matrix}\)\(\begin{matrix} u^{-r}_2 \\ -rv^{-r}_2 \end{matrix}\),\\\nn
&\(\begin{matrix} - am & i \partial_0 - r\(k + 2r aH \xi+  \frac{gaQ}{2}\)  \\  i \partial_0 +r \(k +2 r aH \xi +  \frac{gaQ}{2}\)  & -am\end{matrix} \)\(\begin{matrix} u^r_2 \\ rv^r_2 \end{matrix}\) \\ 
&  \quad\qquad\quad\qquad \quad\qquad \quad\qquad \quad\qquad \quad\qquad =  -\frac{gaQ}{2}\(\begin{matrix} 0 & (1+ r) \\ -(1+ r)  & 0 \end{matrix}\)\(\begin{matrix} u^{-r}_1 \\ -rv^{-r}_1 \end{matrix}\),
\end{align}
where here the subscript indicates the $SU(2)$ component. The system decouples into three disjoint sectors; the $\{u_1^+, v_1^{+}\}$ and $\{u_2^{-}, v_{2}^{-}\}$ modes are two decoupled sectors, while the $\{u_1^-, v_1^{-}\}$ and $\{u_2^{+}, v_{2}^{+}\}$ are coupled together. In what follows, we treat each of these coupled and decoupled sectors separately.

\subsubsection{Decoupled sector: $\{u_1^+, v_1^{+}\}$ and $\{u_2^{-}, v_{2}^{-}\}$}\label{sec:decoupled}
\label{sec:DecSec}

The equations of motion for $\{u_1^+, v_1^{+}\}$ and $\{u_2^{-}, v_{2}^{-}\}$  are decoupled, working in the de Sitter limit $\eta \approx -1/(aH)$ and introducing the variable $x = -k\eta$, we have
\begin{align}\label{eq:decoupledeoms}
& \(\begin{matrix} - \frac{\mu}{x} & -i \partial_x - r\(1+ \frac{1}{x}\(2r\xi + (-1)^a \frac{m_{Q}}{2}\) \) \\  -i \partial_x +r \(1+ \frac{1}{x}\(2r\xi + (-1)^a \frac{m_{Q}}{2}\) \)  & -\frac{\mu}{x}\end{matrix} \)\(\begin{matrix} u^r_a \\ rv^r_a \end{matrix}\)  = 0
\end{align}
where $r = +$ when $a = 1$, and $r = -$ when $a = 2$. We have also introduced the rescaled dimensionless mass parameter
\begin{align}
\mu = \frac{m}{H}.
\end{align}
Rescaling the fields by $U^{r}_a = x^{1/2}u^r_a$, and changing variable to $u = 2ix$, eq.\ \eqref{eq:decoupledeoms} can be written \cite{Adshead:2015kza}
\begin{align}\label{eqn:decoupledeqn}
\partial_u^2 U^r_a+\frac{r}{u}\(\frac{1}{2}-i\(2\xi-\frac{m_Q}{2}\)\)U^r_a +\(-\frac{1}{4}+ \frac{\(2\xi-\frac{m_Q}{2}\)^2+\mu^2+\frac{1}{4}}{u^2} \)U^r_a = 0
\end{align}
Eq.\ \eqref{eqn:decoupledeqn} is solved by the Whittaker functions
\begin{align}
U^{r}_a  = A W_{r\kappa, i\tilde{\mu}}(2ix)+B W_{-r\kappa, i\tilde{\mu}}(-2ix)
\end{align}
with 
\begin{align}
\kappa = \(\frac{1}{2}+i\tilde\kappa\),\quad 
\tilde \kappa = \(\frac{m_Q}{2}-2\xi\), &\quad \tilde{\mu} = \sqrt{\(\frac{m_Q}{2}-2\xi\)^2 + \mu^2}. \label{eqn:kappatildemutildedef}
\end{align}
The function $V^r_a = x^{1/2}r v^r_a$ can be found from eq.\ \eqref{eq:decoupledeoms}, and making use of the identities \cite{Adshead:2015kza}
\begin{align}\label{eq:whitidents}
z\(-\partial_z - \(\frac{1}{2} - \frac{\kappa}{z}\)\)W_{\kappa, \tilde{\mu}}(z) = & -\( \(\frac{1}{2} -\kappa\)^2 - \tilde{\mu}^2\)W_{\kappa-1, \tilde \mu}(z), \\
z\(-\partial_z +\(\frac{1}{2} - \frac{\kappa}{z}\)\)W_{\kappa, \tilde{\mu}}(z) = & W_{\kappa+1, \tilde{\mu}}(z).
\end{align}
Imposing the normalization condition $|u^+_1|^2 + |v^+_1|^2 = |u^-_2|^2 + |v^-_2|^2 = 2$, and demanding that the solutions approach positive frequency modes in the infinite past, $\lim_{x\to\infty} u,v \propto e^{ix}$, we  obtain the solutions
\begin{align}\label{eq:decoupledmassivesols1}
u_{1}^+&  = \frac{\mu e^{{\pi\tilde\kappa}/{2}}}{\sqrt{x}} W_{-\frac12- i \tilde\kappa, i\tilde{\mu}}(-2ix)\quad 
v^+_1   = i \frac{ e^{{\pi\tilde\kappa}/{2}}}{\sqrt{x}} W_{\frac12-i \tilde\kappa,i\tilde{\mu}}(-2ix)
\end{align}
and
\begin{align}\label{eq:decoupledmassivesols2}
u_{2}^-&  = \frac{e^{-{\pi\tilde\kappa}/{2}}}{\sqrt{x}} W_{\frac12+ i \tilde\kappa, i\tilde{\mu}}(-2ix)\quad 
v_{2}^-  = {+}i\mu\frac{e^{-{\pi\tilde\kappa}/{2}}}{\sqrt{x}}  W_{-\frac12+ i \tilde\kappa, i\tilde{\mu}}(-2ix).
\end{align} 
Note that in this sector, the gauge fields and axion always enter in the same way via the parameter $\tilde\kappa$.

\subsubsection{Coupled sector: $\{u_1^-, v_1^{-}\}$ and $\{u_2^{+}, v_{2}^{+}\}$}\label{sec:coupled}
\label{sec:CoupSec}

The equations of motion for the pair $\{u_1^-, v_1^{-}\}$ are coupled to the equations for $\{u_2^{+}, v_{2}^{+}\}$ via the background gauge field
\begin{align}\label{eqn:coupledeoms}
& \(\begin{matrix} - am & i \partial_0 +\({k}-2 aH \xi- \frac{gaQ}{2}\)  \\  i \partial_0 - \(k -2 aH \xi- \frac{gaQ}{2}\)  & -am\end{matrix} \)\(\begin{matrix} u^-_1 \\ -v^-_1 \end{matrix}\)  =  -\frac{gaQ}{2}\(\begin{matrix} 0 & 2 \\ -2  & 0 \end{matrix}\)\(\begin{matrix} u^{+}_2 \\ v^{+}_2 \end{matrix}\),\\\label{eqn:coupledeoms2}
&\(\begin{matrix} - am & i \partial_0 - \(k + 2aH \xi+  \frac{gaQ}{2}\)  \\  i \partial_0 + \(k +2  aH \xi +  \frac{gaQ}{2}\)  & -am\end{matrix} \)\(\begin{matrix} u^+_2 \\ v^+_2 \end{matrix}\)  =  -\frac{gaQ}{2}\(\begin{matrix} 0 & 2 \\ -2  & 0 \end{matrix}\)\(\begin{matrix} u^{-}_1 \\ -v^{-}_1 \end{matrix}\).
\end{align}
In full generality, analytical solutions to these equations are difficult to find; they decouple into fourth order ordinary differential equations. These fourth order equations do not appear to have known closed-form solutions. In the appendices we discuss several analytic approaches to solving the system in eq.\ \eqref{eqn:coupledeoms} and \eqref{eqn:coupledeoms2}. In appendix \ref{app:WKB}, we develop a Wentzel-Kramers-Brillouin (WKB) type analysis to find an asymptotic series solution. In appendix \ref{app:seriessols}, we use an extended Frobenius method to find power-series solutions (in $x= k/aH$) in the regions where $x \to \infty$. Series solutions are also possible in the limit $x \to 0$, but we do not present them here. Finally, in appendix \ref{app:PVsols} we find asymptotic solutions in the large mass limit.

\paragraph{Massless limit:}

In the massless limit, eqs.\ \eqref{eqn:coupledeoms} and \eqref{eqn:coupledeoms2}  reduce to a pair of coupled equations that can be solved analytically. In this limit, the equations of motion can be written 
\begin{align}
\(\begin{matrix}-i \partial_x + \(1 +\frac{1}{x}\(2  \xi+ \frac{m_Q}{2}\)\)   & -\frac{m_Q}{x} \\ -\frac{m_Q}{x} &  -i \partial_x - \(1 - \frac{1}{x}\(2  \xi+ \frac{m_Q}{2}\)\)  \end{matrix} \)\(\begin{matrix} u^+_2 \\ u^{-}_1  \end{matrix}\) = & 0,
\end{align}
and 
\begin{align}
\(\begin{matrix} - i \partial_x +\(1-\frac{1}{x}\(2  \xi+ \frac{m_Q}{2}\)\) &\frac{m_Q}{x}\\ \frac{m_Q}{x} & -i \partial_x - \(1 + \frac{1}{x}\(2  \xi+ \frac{m_Q}{2}\)\) \end{matrix} \)\(\begin{matrix} -v^{-}_1\\ v^+_2   \end{matrix}\)  =  &0.
\end{align}
Rephasing the modefunctions
\begin{align}\label{eq:rephase}
\(\begin{matrix} u^+_2 \\ u^{-}_1  \end{matrix}\) =e^{-i\(2  \xi+ \frac{m_Q}{2}\) \ln x} \(\begin{matrix} \tilde{u}^+_2 \\ \tilde{u}^{-}_1 \end{matrix}\), \quad \(\begin{matrix} \tilde{v}^{-}_1\\ \tilde{v}^+_2   \end{matrix}\) = e^{i\(2  \xi+ \frac{m_Q}{2}\) \ln x}\(\begin{matrix} -v^{-}_1\\ v^+_2   \end{matrix}\),
\end{align}
the equations become
\begin{align}\label{eqn:rephase}
\(\begin{matrix}-i \partial_x +1   & -\frac{m_Q}{x} \\ -\frac{m_Q}{x} &  -i \partial_x -1  \end{matrix} \)\(\begin{matrix} \tilde{u}^+_2 \\ \tilde{u}^{-}_1  \end{matrix}\) = & 0, \quad \(\begin{matrix}-i \partial_x +1   & \frac{m_Q}{x} \\ \frac{m_Q}{x} &  -i \partial_x -1  \end{matrix} \)\(\begin{matrix}\tilde{v}^{-}_1\\ \tilde{v}^+_2   \end{matrix}\) =  0.
\end{align}
Rescaling ${U}^\pm_a  = \sqrt{x}\tilde{u}^\pm_a$, ${V}^\pm_a  = \sqrt{x}\tilde{v}^\pm_a$, and changing variable to $u = 2ix$, these equations can be decoupled into second order equations 
\begin{align}
 \partial^2_u{V}^{-}_1+ \frac{1}{2u}{V}^{-}_1 + \(-\frac{1}{4} +\frac{1+4m_Q^2}{4u^2} \) {V}^{-}_1 =  & 0,\\
\partial^2_uU^-_1 -  \frac{1}{2u}U^-_1 + \(-\frac{1}{4}  +\frac{1+4m_Q^2}{4u^2} \)U^-_1 =  & 0,
\end{align}
which take the form of the Whittaker equation. The solutions for $U^-_1$ and $V^{-}_1$ are the pairs
\begin{align}
U^-_1(x) = & A W_{-\frac{1}{2}, i m_Q} (2 i x)+ B W_{\frac{1}{2}, i m_Q} (-2 i x),\\
V^-_1(x) = & C W_{\frac{1}{2}, i m_Q} (2 i x)+ D W_{-\frac{1}{2}, i m_Q} (-2 i x).
\end{align}
The solutions for $U^+_2(x)$ and $V^+_2(x)$ can be obtained from eq.\ \eqref{eqn:rephase} by making use of the identities in eq.\ \eqref{eq:whitidents}
\begin{align}
U^+_2 =  \frac{x}{m_{Q}}\(-i \partial_x +\frac{i}{2x} -1\) U^{-}_1, \quad V^+_2 = - \frac{x}{m_{Q}}\(-i \partial_x +\frac{i}{2x} +1\) V^{-}_1.
\end{align}
Imposing the normalization condition $|u^-_1|^2 + |u^+_2|^2 = |v^-_1|^2 + |v^+_2|^2 = 2$, and demanding that the solutions approach positive frequency modes in the infinite past, $\lim_{x\to\infty} u,v \propto e^{ix}$, we  obtain the solutions
\begin{align}\nn\label{eq:coupledmasslesssols}
u_{1}^{-} = &  \frac{e^{-i\(2  \xi+ \frac{m_Q}{2}\) \ln x}}{\sqrt{x}} W_{\frac{1}{2}, i m_Q} (-2 i x),\;\;\,\quad\quad \quad
u^{+}_2 = - i  m_{Q} \frac{e^{-i\(2  \xi+ \frac{m_Q}{2}\) \ln x}}{\sqrt{x}}W_{-\frac{1}{2}, i m_Q} (-2 i x),\\
v_1^{-} = & - m_{Q} \frac{e^{i\(2  \xi+ \frac{m_Q}{2}\) \ln x}}{\sqrt{x}}W_{-\frac{1}{2}, i m_Q} (-2 i x), \quad
v_{2}^{+} =  - i\frac{e^{i\(2  \xi+ \frac{m_Q}{2}\) \ln x}}{\sqrt{x}}W_{\frac{1}{2}, i m_Q} (-2ix).
\end{align}
Note that, in this limit, the axion appears in these solutions simply as a phase. As we show below, these cancel when computing currents, and thus the axion has no effect on the coupled fermion solutions in the massless limit.

\section{Currents and backreaction}\label{sec:currentsandbackreact}

Now that we have found and quantized  solutions to the equations of motion in the background axion and gauge fields, we can quantify their effect on the background. In this section, we begin by writing down the backreacted equations of motion, and defining the backreaction currents for the gauge field and axion equations of motion. We then introduce our regularization scheme and show that all divergences can be absorbed by renormalizing the parameters  of the theory. Finally, we compute the regularized and renormalized currents and discuss their backreaction on the equations of motion.

\subsection{Backreacted equations of motion}
The equations of motion for the background, including the effects of the backreaction of the particles, are found from varying the action in eq.\ \eqref{eqn:backgroundaction} with respect to the axion, $\phi$, and the gauge field, $A^a_{\mu}$, yielding \cite{Adshead:2012kp}
\begin{align}\label{axback}
\ddot{\axion}+3H\dot{\axion} + \frac{dV(\axion)}{d\axion} + 3\frac{\alpha}{\pi f}g(HQ^3 + \dot{Q}Q^2) = & \frac{1}{a^3}\frac{\lambda}{f}\nabla_{\mu}(\bar{\psi}\gamma^\mu\gamma^5 \psi) + \frac{1}{a^3}\frac{\delta \Delta\mathcal{L}}{\delta\phi},\\ \label{gaugeback}
\ddot{Q} + 3H \dot{Q}+(\dot{H}+2H^2)Q + 2 g^2 Q^3 - g\frac{\alpha}{\pi f}Q^2 \dot{\axion} =  & -\frac{g\delta_i^a}{3a^3} \bar{\psi} \gamma^i \tau^a \psi +\frac{\delta^{a}_i}{3a^4} \frac{\delta \Delta\mathcal{L}}{\delta A^{a}_i}.
\end{align}
In these expressions, the terms
\begin{align}\nn
\frac{1}{a^3}\frac{\delta \Delta\mathcal{L}}{\delta\phi} = & \frac{\lambda^2}{6 \pi^2 f^2}\[-\Box^2 \phi  - \nabla_{\mu}(G^{\mu\nu}\partial_\nu \phi) + \frac{4\lambda^2}{f^2} \nabla_\mu\(\partial^\mu \phi \partial_\nu \phi\partial^\nu \phi\)\]+ 6\frac{\lambda \alpha}{\pi f}g(HQ^3 + \dot{Q}Q^2),\\
\frac{\delta^{a}_i}{3a^4} \frac{\delta \Delta\mathcal{L}}{\delta A^{a}_i} = & 2 g\lambda\frac{\alpha}{\pi f}Q^2 \dot{\axion},
\end{align}
are the contribution from the terms induced by the change of variable from eq.\ \eqref{eqn:lagrangianshift}.

Making use of the equations of motion for the fermion fields, the terms involving the fermions on the right hand side of eq.\ \eqref{axback} can be written as
\begin{align}
 \frac{\lambda}{f}\nabla_{\mu}(\bar{\psi}\gamma^\mu\gamma^5 \psi) =- i2\mu   \frac{\lambda}{f}(\bar{\psi}\gamma^5 \psi).
\end{align}
Taking the vacuum expectation value (vev), we define \cite{Adshead:2018oaa}\footnote{Note that our results differ significantly from the work of Ref.\  \cite{Mirzagholi:2019jeb}. In particular, we find a non-vanishing backreaction on the axion equations of motion.  We demonstrate in appendix \ref{app:antisym} that the anti-symmetrization procedure advocated in Ref.\ \cite{Mirzagholi:2019jeb} gives the identical result.}
\begin{align}
\mathcal{B} = - 2i m  \frac{\lambda}{f}\langle \bar{\psi}\gamma^5 \psi \rangle,
\end{align}
for the backreaction on the axion equation of motion, and \cite{Mirzagholi:2019jeb}
\begin{align} 
\mathcal{J} =  -\frac{\delta_i^a}{3a^3} \langle \bar{\psi} \gamma^i \tau^a \psi\rangle,
\end{align}
for the backreaction on the gauge field equation of motion.

\subsection{Regularization}

The backreaction currents, $\mathcal{B}$ and $\mathcal{J}$,  are quadratic expectation values of the quantum fermion field. As is usual in quantum field theory, these quantities are divergent and must be regularized.  
Schematically
\begin{align}
\la O \ra = \la \psi^\dag \mathbb O  \psi \ra  = \la \psi^\dag_\alpha \mathbb O_{\alpha\beta}  \psi_\beta \ra .
\end{align}
Using the field operator defined in Eq.\eqref{eq:modeexpansion2}, we can write down the vev in the Bunch Davis vacuum explicitly as 
\be\label{eqn:BDexpectation}
\begin{split}
\la O \ra  & =\bra{ 0} \int d^3x \int \frac{d^3 k }{(2\pi)^{3/2}}  \int \frac{d^3 k' }{(2\pi)^{3/2}}  \sum_j \[ {U}^*_{j,\alpha}(\eta, \bt k')a^\dag_j(\bt k') + {V}^*_{j,\alpha} (\eta, \bt k') b_j(-\bt k') \] \\
&\qquad O_{\alpha\beta}  \[ {U}_{j,\beta}(\eta, \bt k)a_j (\bt k)+ {V}_{j,\beta}(\eta, \bt k) b_j^\dag(-\bt k) \]\ket { 0}e^{i \bt x \dot (\bt k-\bt k')}\\
& = \int \frac{d^3 k }{(2\pi)^3}   \sum_{j=1}^4 \[ {V}^*_{j,\alpha} (\eta, \bt k)O_{\alpha\beta}{V}_{j,\beta}(\eta, \bt k)    \].
\end{split}
\ee
Generically, the integrals appearing in eq.\ \eqref{eqn:BDexpectation} are UV divergent. In order to regularize $\la O \ra$, we introduce a set of massive regulator (Pauli-Villars \cite{Pauli:1949zm}) fields $\psi_n$ with masses $M_n$ and define the regularized currents\footnote{For examples of applications of Pauli-Villars regularization to cosmological correlation functions, see for example, \cite{Weinberg:2010wq, Xue:2012wi}. Our implementation is somewhat different to the methods described there. However, we have checked that the method used here coincides with the results one obtains using dimensional regularization in Minkowski space, as well the result from adiabatic subtraction in de Sitter space \cite{Adshead:2021ezw} in the axion only case. The methods based on the instantaneous diagonalization of the Hamiltonian described by Refs.\ \cite{Figueroa:2013vif, Mirzagholi:2019jeb} do not completely remove the divergent behavior, and do not appear to be readily applicable here.} 
\begin{align}
\mathcal{B}_{\rm reg} =  & - 2 i   \frac{\lambda}{f}\(m\langle \bar{\psi}\gamma^5 \psi \rangle+\sum_n M_n Z_n^{-1}\langle \bar{\psi}_n\gamma^5 \psi_n \rangle\),\\
\mathcal{J}_{\rm reg} = & -\frac{\delta_i^a}{3a^3} \(\langle \bar{\psi} \gamma^i \tau^a \psi\rangle + \sum_n Z_n^{-1} \langle \bar{\psi}_n \gamma^i \tau^a \psi_n \rangle\).
\end{align}
The currents $\mathcal{B}$ and $\mathcal{J}$ are logarithmic and quadratically divergent. To cancel the divergences, we impose the relations among the $Z_n$ and $M_n$, which read
\begin{align}\label{eqn:PauliVillars}
\sum_{n = 1}^N Z_n^{-1} = -1, \quad \sum_{n = 1}^N Z_n^{-1} M_n^2 = -m^2.
\end{align}
Clearly $N = 2$ fields are required to implement these conditions, however, we leave the number of fields arbitrary in what follows.  For notational compactness, we write $\psi = \psi_0$, $Z_0 = 1$ and $M_0 = m$, so that the conditions can be written
\begin{align}\label{eqn:PauliVillars2}
\sum_{n = 0}^N Z_n^{-1} = 0, \quad \sum_{n = 0}^N Z_n^{-1} M_n^2 = 0.
\end{align}
We next split the contributions to the backreaction quantities into terms coming from the decoupled and coupled sectors
\Beq
\mathcal{J}&=\mathcal{J}^{\rm dec}+\mathcal{J}^{\rm coup},\\
\mathcal{B}&=\mathcal{B}^{\rm dec}+\mathcal{B}^{\rm coup},
\Eeq
 and consider each in turn. The reason for the split is because the decoupled sectors can be computed analytically, while the coupled sectors require some numerical evaluation.

\subsection{Decoupled sector: $\{u_1^+, v_1^{+}\}$ and $\{u_2^-, v_2^{-}\}$}

The expressions for the backreaction quantities in the decoupled sector are given by
\Beq
\mathcal{J}^{\rm dec}_{\rm reg} = \sum_{n = 0}\mathcal{J}^{\rm dec}_{n}  =\frac{H^3}{6}\frac{1}{2\pi^2}\int x^3 d\ln x \sum_{n = 0} Z_n^{-1}\(|u_{1,n}^{+}|^2-|v_{1,n}^{+}|^2+|u_{2,n}^{-}|^2-|v_{2,n}^{-}|^2\),
\Eeq
and
\Beq
\mathcal{B}^{\rm dec} = \sum_{n = 0}\mathcal{B}^{\rm dec}_{n} =\frac{\lambda H^4}{\pi^2f}\int x^3 d\ln x \sum_{n = 0} Z^{-1}_n \mu_n \frac{\Im(u_{1,n}^{+\star}v_{1,n}^++v_{2,n}^{-\star}u_{2,n}^-)}{2i}.
\Eeq
After substituting the Bunch-Davies solutions, see eqs.\ \eqref{eq:decoupledmassivesols1}, and \eqref{eq:decoupledmassivesols2} and integrating $x$ from $0$ to a UV-regulator $\Lambda\gg1$ we arrive at 
\begin{align}\nn
 \mathcal J^{\rm dec}_{n}(\Lambda) 
  & =  Z_n^{-1} \frac{ H^3}{12\pi^2}\Bigg[4\mu_n^2   \tilde\kappa \ln (2\Lambda )+4 \gamma  \mu_n ^2  \tilde\kappa -7 \mu_n ^2  \tilde\kappa +\frac{8 \tilde\kappa^3}{3}-\frac{ 4 \tilde\kappa }{3} \\\nn
  &+2 \mu_n ^2  \tilde\kappa  \sum_{r=\pm1}\sum_{b=\pm1} \bigg\{\Re\[H_{i (r  \tilde\kappa + b  \sqrt{\mu_n ^2+  \tilde\kappa ^2})-1}\]+\text{csch}\left(2 \pi  \sqrt{\mu_n ^2+ \tilde\kappa ^2}\right)\\\nn
&\qquad  \times \sinh \left(  -\pi\sqrt{\mu_n ^2+ \tilde\kappa^2}-r\pi  \tilde\kappa  \right) e^{\pi  b \left(\sqrt{\mu_n ^2+ \tilde\kappa ^2}-r \tilde\kappa  \right)} \Re\left[H_{i b \left( r  \tilde\kappa - \sqrt{\mu_n ^2+ \tilde\kappa ^2}\right)+2}\right] \bigg\}\\
  &-2 \text{csch}\left(2 \pi  \sqrt{\mu_n ^2+  \tilde\kappa ^2}\right) \left({C_-} \sinh (2 \pi \tilde\kappa )-{C_+} \sinh \left(2 \pi  \sqrt{\mu_n ^2+ \tilde\kappa ^2}\right)\right)\Bigg],
\end{align}
where $\tilde{\kappa}$ was defined above in eq.\ \eqref{eqn:kappatildemutildedef}, $H_n$ is the harmonic number, and
\begin{align}
&C_-=\left(\frac{1}{3} \left(2 \tilde{\kappa} ^2+\frac{\mu_n ^2}{2}-1\right)-\frac{\left(\mu_n ^2+1\right)^2 \mu_n ^2}{4 \tilde{\kappa} ^2+\left(\mu_n ^2+1\right)^2}-\frac{ \left(\mu_n ^2+4\right)^2\mu_n ^2}{2 \left(16 \tilde{\kappa} ^2+\left(\mu_n ^2+4\right)^2\right)}\right) \sqrt{\tilde{\kappa} ^2+\mu_n ^2},\\
&C_+=  \left( \frac{(\mu_n ^4-16)\mu_n ^2}{2(16 \tilde{\kappa} ^2+\left(\mu_n ^2+4\right)^2)} +\frac{\left(\mu_n ^4-1\right)\mu_n ^2}{4 \tilde{\kappa} ^2+\left(\mu_n ^2+1\right)^2}-\frac{2 \tilde{\kappa} ^2}{3}+\frac{1}{3}\right)\tilde{\kappa}.
\end{align}
The backreaction current on the axion similarly reads
\begin{align}
\mathcal B^{\rm dec}_{n}(\Lambda) &= Z_n^{-1}\mu^2_n\frac{ \lambda H^4}{\pi^2f}\bigg\{  \left[{3 \tilde\kappa } \ln (2\Lambda )  + \tilde\kappa\left(3\gamma_E  -{15/2  } \right)\right]\\\nn
&\quad+ \sum_{r, b=\pm} \bigg\{\frac{1}{2}\Im\left[\left(\mu_n ^2-2 \tilde\kappa ^2-3 i r \tilde\kappa +1\right) H_{i \left(- r\tilde\kappa+b\sqrt{\mu_n ^2+ \tilde\kappa ^2} \right)}\right] \\\nn
&\qquad+\left[e^{\pi\left(  r \tilde\kappa - b \sqrt{\mu_n ^2+ \tilde\kappa ^2}\right)} \sinh \left(\pi  \left( r\tilde\kappa+b\sqrt{\mu_n ^2+ \tilde\kappa ^2} \right)\right) \text{csch}\left(2 \pi b \sqrt{\mu_n ^2+\tilde\kappa ^2}\right)\right]\\\nn
&\qquad\times\left(\frac{3b}{4} \sqrt{\mu_n ^2+ \tilde\kappa ^2}+2 r\tilde\kappa-\frac{1}{2}\Im\left[\left(\mu_n ^2-2 \tilde\kappa ^2-3 i r\tilde\kappa +1\right) H_{i \left(-r \tilde\kappa+b\sqrt{\mu_n ^2+ \tilde\kappa ^2} \right)}\right]  \right) \bigg\} \bigg\}.
\end{align}
For some details of the computation of these expressions see appendix \ref{app:analints} and Ref.\ \cite{Adshead:2018oaa}. Note that the dependence on the UV cutoff scale cancels once the conditions in eq.\ \eqref{eqn:PauliVillars2} are imposed. Next, we expand in the limit $\mu_n  = M_n/H\to \infty$ to obtain the regularized currents
\begin{align}\nn
 \mathcal{J}^{\rm dec}_{\rm reg}
  & = \frac{ H^3}{12\pi^2} \Bigg[4 \gamma_E  \mu ^2  \tilde\kappa -7 \mu ^2  \tilde\kappa +\frac{8 \tilde\kappa^3}{3}-\frac{ 4 \tilde\kappa }{3}-\frac{1}{3}\tilde{ \kappa}  \left(4 \tilde{\kappa} ^2-12 \mu ^2-3\right) \\\nn
  &+2 \mu ^2  \tilde\kappa  \sum_{r=\pm1}\sum_{b=\pm1} \bigg\{\Re\[H_{i (r  \tilde\kappa + b  \sqrt{\mu ^2+  \tilde\kappa ^2})-1}\]+\text{csch}\left(2 \pi  \sqrt{\mu ^2+ \tilde\kappa ^2}\right)\\\nn
&\qquad  \times \sinh \left(  -\pi\sqrt{\mu ^2+ \tilde\kappa^2}-r\pi  \tilde\kappa  \right) e^{\pi  b \left(\sqrt{\mu ^2+ \tilde\kappa ^2}-r \tilde\kappa  \right)} \Re\left[H_{i b \left( r  \tilde\kappa - \sqrt{\mu ^2+ \tilde\kappa ^2}\right)+2}\right] \bigg\}\\\nn
  &-2 \text{csch}\left(2 \pi  \sqrt{\mu ^2+  \tilde\kappa ^2}\right) \left({C_-} \sinh (2 \pi \tilde\kappa )-{C_+} \sinh \left(2 \pi  \sqrt{\mu ^2+ \tilde\kappa ^2}\right)\right)\Bigg]\\
&    -  \sum_n Z_n^{-1}\frac{ H^3}{3\pi^2} \tilde{ \kappa}  \mu_n ^2 \log (\mu_n ),
\end{align}
and 
\begin{align}\nn
\mathcal B^{\rm dec}_{\rm reg} &=  \mu^2 \frac{\lambda}{\pi^2 f}H^4 \bigg\{  \left[ \frac{\tilde\kappa}{2}\left(3\gamma_E  -\frac{15}{2  } \right)- \frac{ \tilde{\kappa}  \left(4  \tilde{\kappa} ^2-16 \mu ^2-3\right)}{4 \mu^2}\right]\\\nn
&\quad+ \sum_{r, b=\pm} \bigg\{\frac{1}{2}\Im\left[\left(\mu ^2-2 \tilde\kappa ^2-3 i r \tilde\kappa +1\right) H_{i \left(- r\tilde\kappa+b\sqrt{\mu ^2+ \tilde\kappa ^2} \right)}\right] \\\nn
&\qquad+\left[e^{\pi\left(  r \tilde\kappa - b \sqrt{\mu ^2+ \tilde\kappa ^2}\right)} \sinh \left(\pi  \left( r\tilde\kappa+b\sqrt{\mu ^2+ \tilde\kappa ^2} \right)\right) \text{csch}\left(2 \pi b \sqrt{\mu ^2+\tilde\kappa ^2}\right)\right]\\\nn
&\qquad\times\left(\frac{3b}{4} \sqrt{\mu ^2+ \tilde\kappa ^2}+2 r\tilde\kappa-\frac{1}{2}\Im\left[\left(\mu ^2-2 \tilde\kappa ^2-3 i r\tilde\kappa +1\right) H_{i \left(-r \tilde\kappa+b\sqrt{\mu ^2+ \tilde\kappa ^2} \right)}\right]  \right) \bigg\} \bigg\}\\
&- \sum_n  Z_n^{-1} \frac{\lambda}{\pi^2 f}H^4\[-3 \tilde{\kappa} \mu_n ^2 \log (\mu_n )\],
\end{align}
where $\gamma_E$ is the Euler-Mascheroni constant. Notice that both of these regularized expressions still depend on the Pauli-Villars masses and diverge in the limit $\mu_n \to \infty$.  We demonstrate below that when combined with the contributions from the coupled sector,  this dependence on the Pauli-Villars masses either cancels in the gauge current case, or can  be absorbed by renormalizing the axion wavefunction in the axion backreaction case. Note also the appearance of the mass-independent terms in $\mathcal B^{\rm dec}_{\rm reg}$. As we demonstrate below these are due to the usual axial anomaly \cite{Adler:1969gk, Bell:1969ts} from the gauge field as well as the background axion \cite{Adshead:2021ezw}. In the massless limit, these terms (combined with similar terms from the coupled sector, see below) precisely cancel those induced by the field redefinition in eq.\ \eqref{eqn:lagrangianshift} as required by chiral symmetry.

\subsection{Coupled sector: $\{u_1^-, v_1^{-}\}$ and $\{u_2^{+}, v_{2}^{+}\}$}

The backreaction quantities in the coupled sector are given by
\Beq
\mathcal{J}^{\rm coup}_{\rm reg}&=\frac{ H^3}{6}\frac{1}{2\pi^2}\int x^3 d\ln x \sum_{n = 0}Z_n^{-1}\sum_{m=2}^{3}\Bigg[|v_{1m,n}^{-}|^2+|v_{2m,n}^{+}|^2-|u_{1m,n}^{-}|^2-|u_{2m,n}^{+}|^2\\ &  \qquad \qquad\qquad \qquad\qquad \qquad+4\Re(u_{1m,n}^{-\star}u_{2m,n}^{+}+v_{1m,n}^{-\star}v_{2m,n}^{+})\Bigg]\\
&\equiv\frac{ H^3}{6\pi^2}\int x^3 d\ln x  \sum_{n = 0}\mathcal{P}_{\mathcal{J},n}^{\rm coup}(x),
\Eeq
and
\Beq
\mathcal{B}^{\rm coup}_{reg}&=\frac{\lambda H^4}{\pi^2f}\int x^3 d\ln x \sum_{n=0}Z_n^{-1}\mu_n \sum_{m=2}^{3}\frac{\Im(u_{1m,n}^{-\star}v_{1m,n}^-+v_{2m,n}^{+\star}u_{2m,n}^+)}{2i}\\
&\equiv\frac{\lambda H^4}{\pi^2f}\int x^3 d\ln x\sum_{n=0} Z_n^{-1} \mathcal{P}_{\mathcal{B},n}^{\rm coup}(x).
\Eeq
The general Bunch-Davies solutions in the coupled sector can be found only numerically, so we cannot provide analytic expressions for the UV-regulated backreaction quantities.  To proceed we split the integration over $\mathcal{P}$ and write each case as
\begin{align}\label{eqn:PVsplit}
\lim_{\Lambda \to \infty} \int_0^\Lambda x^3 d\ln x  \sum_{n = 0}\mathcal{P}_{\mathcal{J},n}^{\rm coup}(x) = \int_0^q x^3 d\ln x  \sum_{n = 0}\mathcal{P}_{\mathcal{J},n}^{\rm coup}(x)+\lim_{\Lambda \to \infty} \int_q^\Lambda x^3 d\ln x  \sum_{n = 0}\mathcal{P}_{\mathcal{J},n}^{\rm coup}(x),
\end{align}
where $q$ is an intermediate scale, $\mu \ll q \ll \mu_n$, and the UV scale $\mu_n \ll \Lambda$. As $\mu_n \to \infty$, the regulator fields give a vanishing contribution to the first integral, and we may ignore them \cite{Weinberg:2010wq}. We focus our attention on the second term.

In the region $\mu, m_Q, \xi \ll q$ and $ q \gg 1$ it is possible to find a series solution to the coupled equations of motion (see appendix \ref{app:seriessols}). Using these series solutions, we can then find the following series representations of the backreaction currents for the physical fields 
\Beq\label{eqn:physJ}
\mathcal{P}_{\mathcal{J}, 0}^{\rm coup}(x) 
  & = 8 m_Q x^{-1}+2\left[\mu ^2 \left(4\xi - m_Q\right)-2m_Q\left( m_Q^2+1\right)\right]x^{-3}\\
&+\Bigg[4 \mu ^2\xi \left(16\xi ^2- 3
   \mu ^2-6 m_Q^2-5\right)\\
&+2m_Q \left(5 \mu ^2+6+m_Q^2\frac{\left(\mu
   ^2+15\right) +3 m_Q^2}{2}\right)\Bigg]
   x^{-5}+\mathcal{O}\left(x^{-7}\right),
\Eeq
and
\Beq\label{eqn:physB}
\mathcal{P}_{\mathcal{B}, 0}^{\rm coup}(x) 
  & =\mu^2  \left(m_Q+4 \xi \right)\left[\frac{3}{2}x^{-3}+\frac{5}{4}\left(8 \xi  m_Q-3 \mu ^2-2
   m_Q^2+16 \xi ^2-5\right) x^{-5}\right]+\mathcal{O}\left(x^{-7}\right).
\Eeq
Unfortunately, these series solutions are inaccurate in the regions where $x \sim \mu_n$ and $x < \mu_n$, and thus we require a different approximation scheme to find the contributions from the regulator fields. In the region $\mu_n \gg 1$, the solutions can be accurately approximated using the WKB expansion (see appendices \ref{app:WKB} and \ref{app:PVsols}). Using these large mass WKB solutions, and working in the limit $\mu_n \to \infty$, we find for the regulator fields
\begin{align}\nn\label{eqn:regJ}
& \mathcal{P}_{\mathcal{J}, n > 0}^{\rm coup}(x) \\ \nn& = Z_n^{-1}\Bigg(\frac{8 m_Q}{\left(x^2+\mu_n^2\right)^{1/2}}+ 2\frac{ (m_Q+4\xi) \mu_n^2 -2 m_Q^3 }{\left(x^2+\mu_n^2\right)^{3/2}}-3(m_Q+4\xi)\frac{ \mu_n^2  m_Q (m_Q+4\xi)+\mu_n^2m_Q
   ^2  }{\left(x^2+\mu_n^2\right)^{5/2}}\\ \nn & 
-\frac{\mu_n^2\left(\mu_n^2-4 x^2\right)}{\left(x^2+\mu_n^2\right)^{7/2}}\frac{(m_Q+4\xi)^3}{4}-\frac{x^2 \left(5 \left(4 x^2-3\mu_n^2\right)\mu_n^2 (m_Q+4\xi) +4 \left(4 x^4+3 x^2\mu_n^2-\mu_n^4\right)
   m_Q \right)}{4 \left(x^2+\mu_n^2\right)^{9/2}}\Bigg)\\
   &  +\ldots,
\end{align}
and
\begin{align}\nn\label{eqn:regB}
\mathcal{P}_{\mathcal{B}, n >0}^{\rm coup}(x) = &  \frac{1}{2}\(m_Q + 4\xi\)   \mu_n^2 Z_n^{-1}\Bigg(\frac{  3x^2}{\left(x^2+1\right)^{5/2}} - \frac{15 m_Q ^2 x^2}{2
   \left(x^2+  \mu_n^2\right)^{7/2}}\\\nn
 & - \frac{\left(5x^2 \left( \left(20
   x^4-37 x^2  \mu_n^2+6  \mu_n^4\right)-\(m_Q + 4\xi\)^2 \left(4 x^4+x^2  \mu_n^2-3  \mu_n^4\right)\right)\right)}{8
   \left(x^2+  \mu_n^2\right)^{11/2}}\Bigg)\\
   & +\ldots,
\end{align}
where the `$\ldots$' indicate terms that are higher order in powers of $1/\sqrt{x^2+  \mu_n^2}$.

Using eqs.\ \eqref{eqn:physJ}-\eqref{eqn:regB}, we integrate the UV parts of the currents to obtain
\begin{align}\nn
& \lim_{\Lambda \to \infty} \int_q^\Lambda x^3 d\ln x  \sum_{n = 0}\mathcal{P}_{\mathcal{J},n}^{\rm coup}(x) \\\nn
= &  - 2\(2m_Q q^2 - \(2(m_Q^3+m_Q) +\mu^2\(\frac{m_Q}{2}-2\xi\)\) \log(2q)\)
\\\nn
&-2\Bigg[ - 2 \(\frac{m_Q}{2}+2\xi\)^2 m_Q+ \frac{2 \(\frac{m_Q}{2}+2\xi\) ^3}{3}-\(\frac{m_Q}{2}+2\xi\)  \left(m_Q ^2+\frac{1}{2}\right)\\\nn
&+m_Q  \left(2m_Q ^2 +\frac{19}{6}\right) -4\xi \mu^2\Bigg]\\
   & + Z_n^{-1}\(4 \mu_n^2 \(\frac{m_Q}{2}-2\xi\)  \log (\mu_n)+4 \left(m_Q  \left(m_Q ^2+1\right) \log (\mu_n)\right)\),
\end{align}
and
\begin{align}\nn
\lim_{\Lambda \to \infty} \int_q^\Lambda x^3 d\ln x  \sum_{n = 0}\mathcal{P}_{\mathcal{B},n}^{\rm coup}(x)  = & - \frac{3 \mu^2 }{2}(m_Q+2\xi)  \log(2q)+ 2\mu^2(m_Q + 2\xi) \\\nn
&+(m_Q + 2\xi)\frac{    \left(3-(m_Q + 2\xi) ^2+6 m_Q^2\right)}{8}\\
& -\sum_{n}Z_n^{-1}\(\frac{3}{2}(m_Q + 2\xi)\mu_n^2 \log\(\mu_n\)\).
\end{align}
Note that the dependence on the UV cutoff, $\Lambda$, has cancelled out, as expected. Further, note that although these expression appear to depend on scale at which we split our integration, $q$, the construction above ensures that the result is independent of this scale once the sum in eq.\ \eqref{eqn:PVsplit} is computed. Similarly to the case in the decoupled sector, we have found terms which depend on the regulator masses, as well as terms that are mass-independent. 

\subsection{Renormalization}\label{sec:renorm}

We have successfully regularized the current, and it remains to absorb the dependence of the Pauli-Villars masses into a redefinition of the parameters of the theory.  Within the approximation we are working, $\dot{\phi}  = $ const., $\ddot{Q}\approx 0$, $\dot{Q} \approx 0$, $\dot{H} \approx 0$,  the equations of motion, eq.\ \eqref{axback}, can be written
\begin{align}\label{axbackapprenorm}
Z_{\phi, \rm bare} (3H\dot{\axion}) +   \frac{dV(\axion)}{d\axion}  + 3(1-2\lambda)\frac{H^4 }{4\pi^2 f} m_Q^3 & -\frac{1}{2\pi^2}\frac{\lambda^2 \dot{\phi}}{Hf^2}\left[3-4\left(\frac{\lambda\dot{\phi}}{fH}\right)^2\right]H^4\\ \nn  & =  - 12 \xi   \frac{\lambda}{\pi^2 f}H^2 \sum_{n=0}Z_n^{-1}M_n^2 \log\(2\frac{M_n}{H}\)+2\mathcal{B}_{\rm ren.} ,\\ \label{gaugebackapprenorm}
\frac{1}{g^2_{\rm bare}}H^3m_Q \(2 + 2 m_Q^2\) - (1-2\lambda)\frac{H^2}{4\pi^2 f}m_Q^2 \dot{\axion}  &=   \mathcal{J}_{\rm ren.} - \frac{H^3}{6\pi^2}  m_Q  \left(2+2m_Q^2\right) \sum_{n=0}  Z_n^{-1} \log \(2\frac{M_n}{H}\),
\end{align}
where we have included a factor $Z_{\phi, \rm bare}$,  the bare coefficient of the axion kinetic term, and we have relabeled the gauge coupling. We have also defined the finite parts of the backreaction currents $\mathcal{B}_{\rm ren.}$ and $\mathcal{J}_{\rm ren.}$.

Note that, in agreement with the results of Ref.\ \cite{Domcke:2018gfr}, the divergent terms in the gauge field equation of motion, eq.\ \eqref{gaugebackapprenorm}, can be absorbed by working with the renormalized gauge coupling $\bar{g}$
\begin{align}\label{eqn:gaugerenorm}
\frac{1}{\bar{g}^2} = \frac{1}{g^2_{\rm bare}}- \frac{2 N_f}{3}\frac{1}{16\pi^2} \sum_{n=0}  Z_n^{-1} \log \(2\frac{H}{M_n}\),
\end{align}
where $N_f = 4$ is the number of Weyl fermions.   The divergent terms in the axion equation of motion, eq.\ \eqref{axbackapprenorm}, can be absorbed into a renormalization of the axion kinetic term $Z_\phi$\footnote{The vanishing of the sum, $\sum_{n = 0}Z_n^{-1}M_n^2 \log\(2\frac{H}{M_n}\) = 0$, is sometimes imposed as an auxiliary condition in Pauli-Villars regularization. Here, we can apparently absorb this divergence into renormalization of the axion kinetic term.}
\begin{align}
\bar{Z}_{\phi} = Z_{\phi, \rm bare}-2\frac{\lambda^2}{\pi^2 f^2} \sum_{n = 0}Z_n^{-1}M_n^2 \log\(2\frac{H}{M_n}\) .
\end{align}
In renormalizing the theory, we have explicitly included the physical fermion in the sum over logs. Had we not, the expressions for the currents would increase without bound as the physical fermion masses are increased. The principle of decoupling indicates that this behavior is unphysical. A minor difficulty then appears in eq.\ \eqref{eqn:gaugerenorm} as the physical fermion becomes massless. In this limit, the physical mass should not be included in the sum. As we see below, in this limit one should include the $\ln(m_Q)$ instead.

\subsection{Massless, or near massless limit and the anomaly}

In the massless limit, we can compute the backreaction currents, $\mathcal{B}$ and $\mathcal{J}$ exactly, analytically.  In this limit, the contribution of the physical fields to the finite or renormalized axion backreaction current, $\mathcal{B}_{\rm ren.}$ vanishes, and the the non-zero part of $\mathcal{B}_{\rm ren.}$ arises from the regulator fields. 

From above we find
\begin{align}\nn
\lim_{\mu \to 0} \mathcal B_{ \rm ren.} &=  \frac{\lambda}{8\pi^2 f}H^4 \bigg\{ - ( m_Q - 4\xi) \left(  (m_Q - 2\xi)^2-3\right)+(m_Q + 4\xi)    \left(3-(m_Q + 4\xi) ^2+6 m_Q^2\right)\bigg\} \\
& =  \frac{\lambda}{\pi^2 f}H^4 \bigg\{ 16\xi^3-3\xi - \frac{3}{4}m_Q^3\bigg\} ,
\end{align}
where in the first line, the first and second terms come from the decoupled and coupled sectors, respectively.

We can verify that the anomaly equation \cite{Adshead:2021ezw} is satisfied. For each Dirac fermion, this reads
\Beq\label{eqn:anomaly}
\langle \nabla_\mu j^{\mu}_5(x)\rangle=-&\frac{\lambda}{f}\frac{\Box^2 \phi}{12\pi^2}-\frac{\lambda}{f}\frac{\nabla_\mu\(G^{\mu\nu}\partial_\nu\phi\)}{12\pi^2}+\frac{1}{3\pi^2}\(\frac{\lambda}{f}\)^3\nabla_\mu \left(\partial^\mu\phi \partial_\nu\phi\partial^\nu\phi\right)\\  -&\frac{\alpha}{4\pi}F_{\mu\nu}\tilde{F}^{\mu\nu}+\frac{1}{384\pi^2}\epsilon^{\mu\nu\alpha\beta}R^{\rho\sigma}{}_{\mu\nu}R_{\rho\sigma\alpha\beta}\,.
\Eeq
Note that in the limit we are working, we can evaluate
\Beq
\Gamma^0{}_{ij}=Ha^2\delta_{ij}\,,\quad G^{00}=3H^2\, ,\quad  \Box\phi = 3H\dot{\phi}, \quad \Box^2\phi = 0, 
\Eeq
and
\Beq
F_{0i}^a = a^2 H Q \delta^a_i, \quad F^a_{ij} = ga^2 Q^2 \epsilon^a_{ij}.
\Eeq
Inserting these into the result from  eq.\ \eqref{eqn:anomaly} above, 
\Beq
\langle \nabla_\mu j^{\mu}_{5}(x)\rangle&=-\frac{1}{4\pi^2}\frac{\lambda \dot{\phi}}{Hf}\left[3-4\left(\frac{\lambda\dot{\phi}}{fH}\right)^2\right]H^4  - \frac{3}{8\pi^2} g^3 H Q^3 =\frac{1}{2\pi^2} \left[16\xi^3 - 3\xi-\frac{3}{4}m_Q^3\right]H^4.
\Eeq
In terms of the backreaction, the anomaly equation is $\mathcal{B} = \frac{\lambda}{f}\nabla_\mu j_5^\mu$, and thus accounting for the fermion doublet which gives an additional factor of 2, we match the anomaly equation.

We can similarly compute the gauge current. This reads
\begin{align}
\lim_{\mu \to 0}\mathcal{J}_{\rm ren.} =\frac{ H^3}{6}\frac{1}{2\pi^2}\int x^3 d\ln x \sum_n \[  ( |{u_{1,n}^+}|^2 - |{v_{1,n}^+}|^2) + ( |{u_{2,n}^-}|^2 - |{v_{2,n}^-}|^2) + 8 \Re{(v_{1,n}^{- *} v_{2,n}^{+}  )}  \]. 
\end{align}

Inserting the solutions from eqs.\ \eqref{eq:decoupledmassivesols1}, \eqref{eq:decoupledmassivesols1}, and   \eqref{eq:coupledmasslesssols}, and regulating the integral with a hard cutoff at $x = \Lambda$, the integrals can be performed analytically using the methods in, for example, Refs.\ \cite{Adshead:2018oaa, Adshead:2019aac}, to obtain
\begin{align}\nn
\lim_{\mu \to 0}\mathcal{J}_{\rm ren.} =\frac{ H^3}{12 \pi^2}\Bigg[  %
  &4 \left(m_Q^3+m_Q\right) \left(\Re\left[H_{-i \sqrt{m_Q^2}}\right]-\gamma_E\right)+2m_Q^3+6m_Q \\ & \qquad \qquad \qquad  -\frac{1}{3}m_Q(16+7m_Q^2 - 36 m_Q \xi)   \Bigg],
\end{align}
where $\gamma_E$ is the Euler-Mascheroni constant, and $H_n$ is the harmonic number. In this expression, the terms on the first line arise from the physical fermions (the coupled modes), while the last term comes from the regulator fields. Note that only the coupled modes give rise to a non-zero contribution from the physical fields.

For large values of $m_Q$, we can expand the harmonic number
\begin{align}
\left(\Re\left[H_{-i \sqrt{m_Q^2}}\right]-\gamma_E\right) = \ln(m_Q) +\frac{1}{12m_Q^2} +\frac{1}{120m_Q^4}+\mathcal{O}(m_Q^{-6}).
\end{align}
Note that the term in $\log(m_Q)$ should be absorbed into the renormalization of the gauge coupling, as noted in Ref.\ \cite{Domcke:2018gfr}. The gauge coupling should then be evaluated at the scale $m_Q$ in order that perturbation theory is under control at large $m_Q$. The remaining terms are consistent with the results found in Ref.\ \cite{Domcke:2018gfr}, who neglect order unity coefficients in their estimates. However, note that the coefficient of the linear term in $m_Q$ appears to have the opposite sign compared with Ref.\ \cite{Domcke:2018gfr}.  The backreaction is vanishing as $m_Q \to 0$.

We can plug these results back into the equations of motion to find, 
\begin{align}\label{axbackapprenorm}
\bar{Z}_{\phi} (3H\dot{\axion}) + \frac{dV(\axion)}{d\axion} + 3\frac{H^4 }{4\pi^2 f}m_Q^3 = &0, \\
\label{gaugebackapprenorm}
\frac{1}{\bar{g}^2}H^3m_Q \(2 + 2 m_Q^2\) - \frac{H^2}{4\pi^2 f}m_Q^2 \dot{\axion} =  &\frac{ H^3}{12 \pi^2}\Bigg[m_Q -\frac{1}{3}m_Q^3\Bigg].
\end{align}
Notice that the effect of the axion-fermion coupling completely cancels out, and the backreaction on the axion equation is vanishing. This is a manifestation of the chiral symmetry, which is restored in the massless limit $\mu = m/H \to0$. Finally, we see that provided the gauge theory does not run to strong coupling at the scale $m_Q$, where $\alpha(m_Q) = 1$, the  backreaction on the gauge field equation of motion is negligible in agreement with the results of Ref.\ \cite{Domcke:2018gfr}.

\subsection{Massive limit and backreaction currents}

We are now ready to compute the backreaction currents away from the massless limit.  While we have analytic solutions for all masses for the decoupled sector, the coupled sector is considerably more complicated. To compute the contributions to the current from the coupled sector, we numerically solve the equations of motion for the fermions and numerically integrate over the solutions. In practice, the naive numerical solutions, obtained from e.g.\ {\sc Mathematica}, are noisy. Because the leading contributions largely cancel, these naive numerical solutions cannot be relied upon to find accurate numerical representations for the backreaction currents. The series solutions obtain in appendix \ref{app:seriessols}, however, are excellent until $x = -k \tau$ becomes comparable to one of the parameters, $x\sim \mu, m_Q, \xi$. Therefore, our strategy is to use the series solutions for as long as possible, then,  while the series solutions are still accurate, we use initialize the numerical solver for the equations of motion, and subsequently numerically integrate over the resulting solutions to find the remaining contributions to the current. The result does not depend on when this transition from series solutions to numerical solutions is made, provided it occurs while the series solutions are still accurate. In practice, to generate the results in what follows, we transition at  $x\sim 200-300$.

The integrals are regulated by UV divergent terms from the integrals over the Pauli-Villars fields, and the results are finite. By construction, the results for the current decrease as the mass increases, and vanish approximately exponentially in $\exp(-\pi \mu)$. This is consistent with naive expectations that particle production of massive particles becomes inefficient for particles with masses larger than the Hubble scale.  

\begin{figure}[t]
  \centering
  \includegraphics[width = \textwidth]{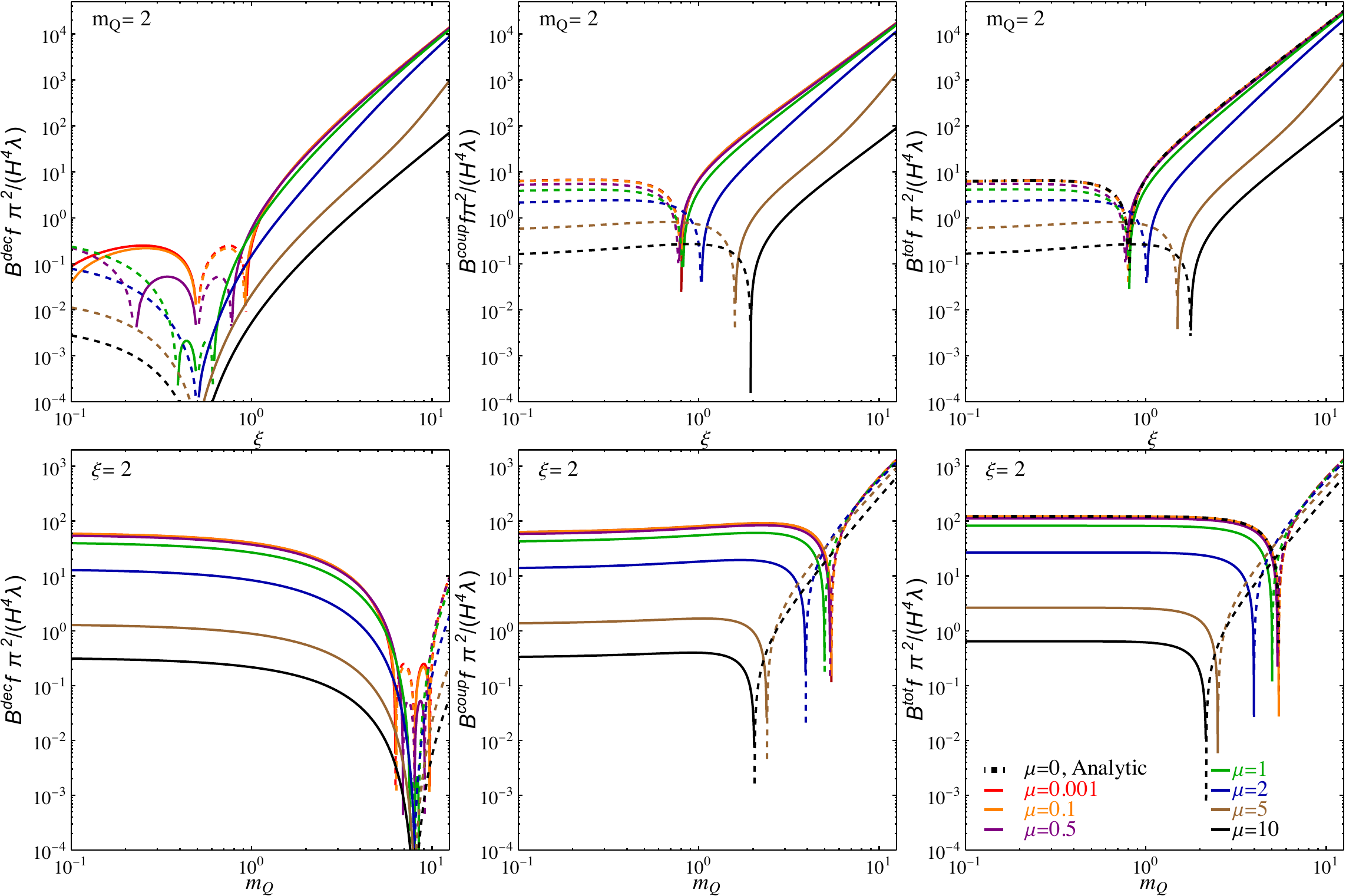}
\caption{Finite or renormalized axial backreaction current, $\mathcal{B}_{\rm ren.}$,  as a function of $m_Q$ at fixed $\xi = 2$ (top panels) and as a function of $\xi$ at fixed $m_Q$ (lower panels) for various fermion masses. The left panels show the contribution to $\mathcal{B}_{\rm ren.}$ due to the decoupled modes, the center panels show the contribution of the coupled modes, and the right panels show their sum. We also show the analytic result in the massless limit in the rightmost panel. Solid lines indicate regions where a quantity is positive, and dashed lines indicate where it is negative. }
\label{fig:AxBackAnom}
\end{figure}

In figure \ref{fig:AxBackAnom} we show the renormalized backreaction current on the axion equation of motion, $\mathcal{B}_{\rm ren.}$.  Note modes from both the coupled and decoupled sectors appear to give equally important contributions to the backreaction in the limit of small $m_Q$ and fixed $\xi$ (top panels), while the coupled sector dominates the current at large values of $m_Q$. At low values of $\xi$ at fixed $m_Q = 2$, the axial backreaction current, $\mathcal{B}_{\rm ren.}$, is dominated by the coupled sector (lower panel). At large $\xi$ both sectors give comparable contributions.  Note that we accurately match on to the analytic result expected from the anomaly equation in the massless limit in all cases.

\begin{figure}[t]
  \centering
  \includegraphics[width = \textwidth]{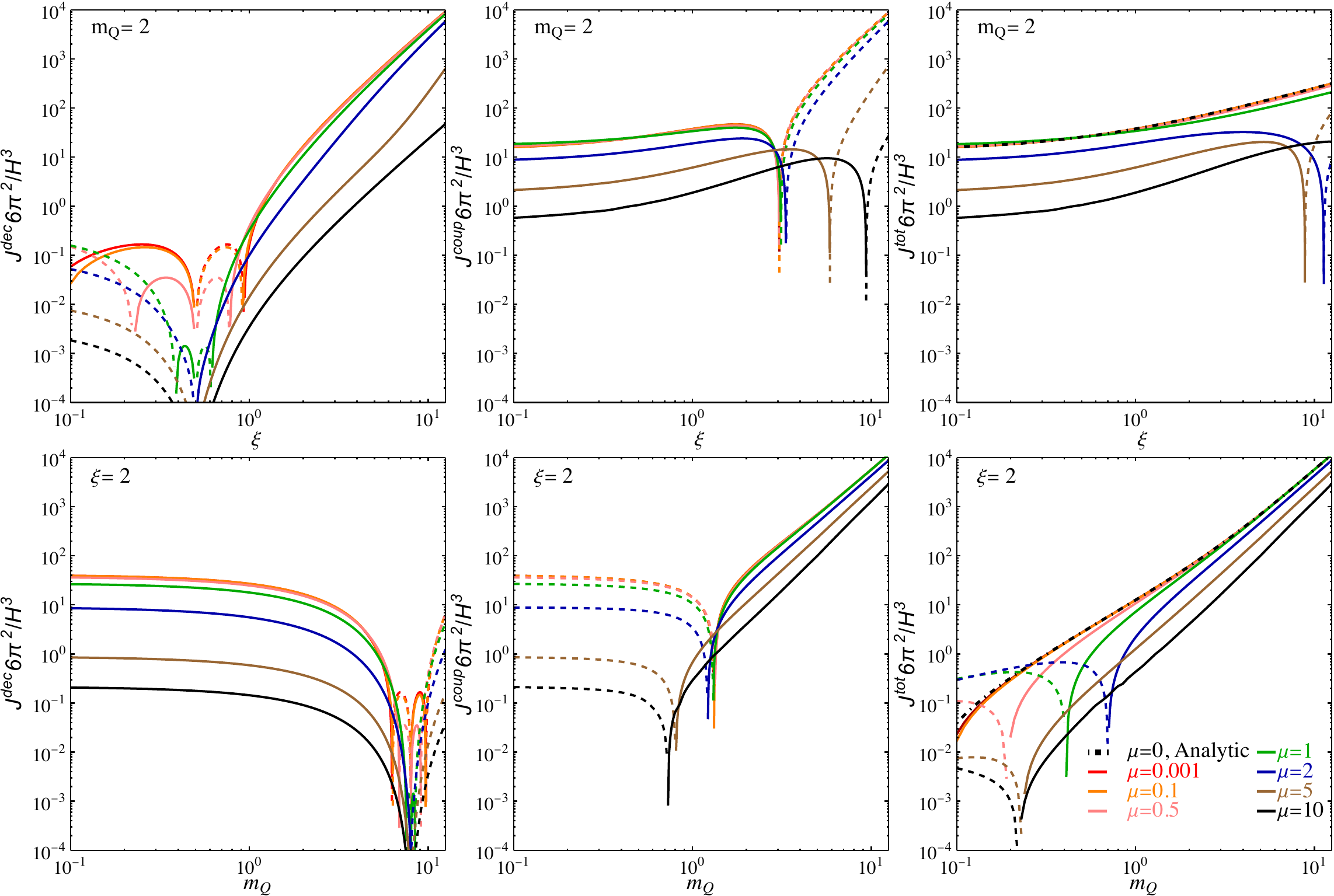}
\caption{Finite or renormalized gauge backreaction current, $\mathcal{J}_{\rm ren.}$, as a function of $m_Q$ at fixed $\xi = 2$ (top panels) and as a function of $\xi$ at fixed $m_Q$ (lower panels) for various fermion masses. The left panels show the contribution to $\mathcal{J}_{\rm ren.}$ due to the decoupled modes, the center panels show the contribution of the coupled modes, and the right panels show their sum. We also show the exact analytic result from the massless limit in the rightmost panel. Solid lines indicate regions where a quantity is positive, and dashed lines indicate where it is negative. }
\label{fig:Gaugeback}
\end{figure}

In figure \ref{fig:Gaugeback} we  show the renormalized backreaction current, $\mathcal{J}_{\rm ren.}$, on the gauge field equation of motion. Here, again, we see the importance of the accurate treatment of the coupled sector. At fixed values of $m_Q$, note that for large $\xi$ there is strong cancellation between the decoupled and coupled contributions to the current (top panels). While at fixed $\xi$ similar behavior can be seen at low $m_Q$, where the contributions to the current are almost exactly cancelling. In fact, these must cancel in the limit the gauge field vanishes, otherwise the fermion backreaction would source a gauge field background representing an instability. Note that our results match accurately onto the analytic result in the massless limit. In generating this curve, we did not subtract off the terms in $\ln(m_Q)$ as described above.

We conclude that accurate results in this model require the accurate treatment of the coupled sector. In most areas of parameter space we either observe strong cancelations between the sectors, or we observe the dominance of the coupled sector. Unfortunately this means that the cumbersome numerical procedures detailed above are required to explore the full phenomenology of the backreaction and particle production in this model. Conclusions based on the analytic results from the decoupled sector alone are unfortunately unreliable.

 \begin{figure}[t]
  \centering
  \includegraphics[width = \textwidth]{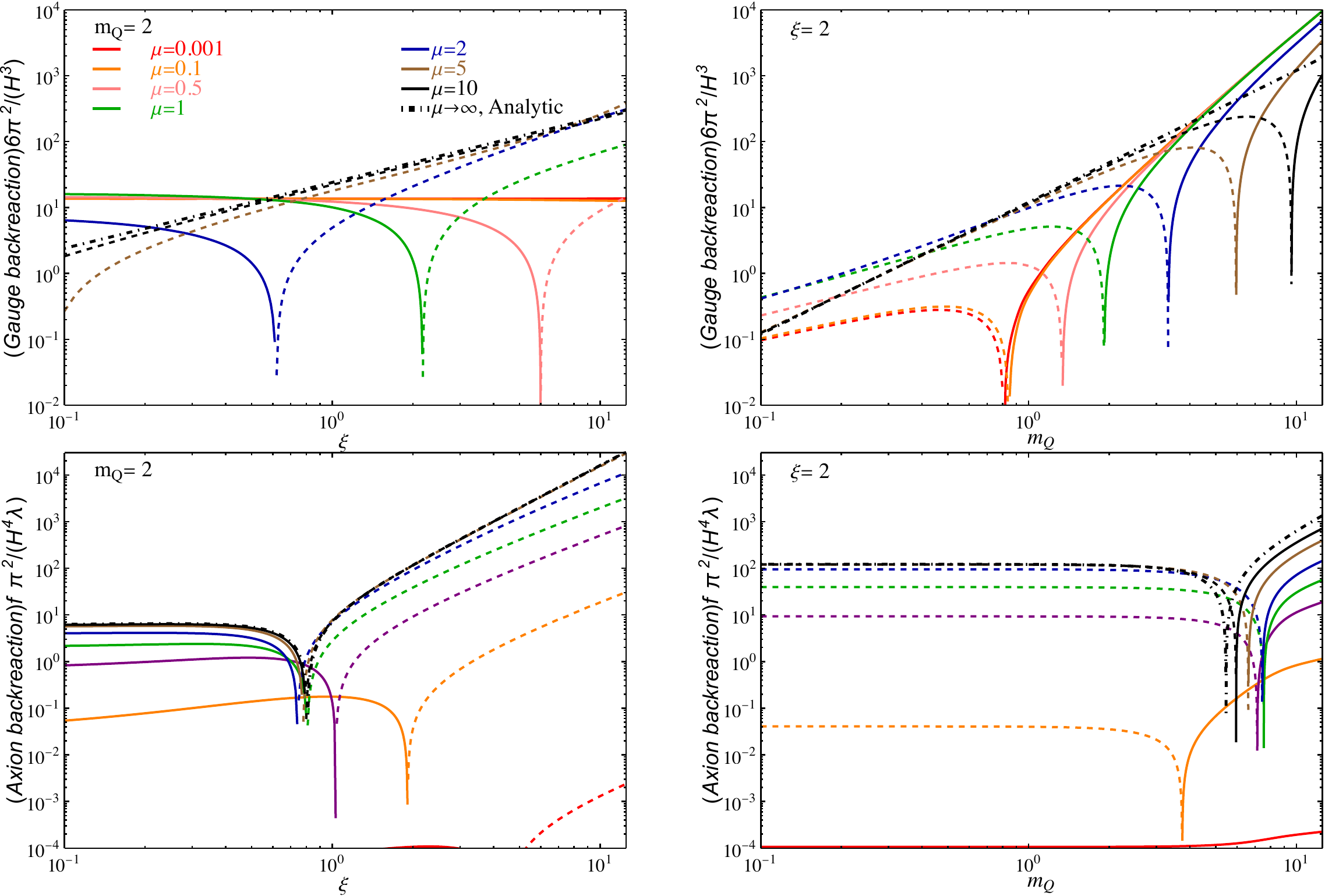}
\caption{We show the backreaction on the  gauge field and axion equations of motion  due to the combination of the renormalized currents and the additional terms induced by the change of variables (the right hand sides of eqns.\ \eqref{gaugeback} and \eqref{axback}, renormalized according to section \ref{sec:renorm}). In the top panels, we display the backreaction to the gauge currents, and in the lower panels we display the axion backreaction. Solid lines indicate regions where the quantity is positive, and dashed indicate where it is negative. We also show the analytic limit $\mu \to \infty$ (black dot-dashed line) where $\mathcal{B}_{\rm ren.} \to 0$ and $\mathcal{J}_{\rm ren.}\to 0$, where the contribution is solely due to that from $\Delta\mathcal{L}$.}
\label{fig:Fullback}
\end{figure}

In figure \ref{fig:Fullback}, we show the right hand sides of the gauge field and axion equations of motion,  eqs.\ \eqref{gaugeback} and \eqref{axback}. These are the terms induced by the fermion interactions, and represent the backreaction on the equations of motion describing the background. In the limit that the fermion masses vanish, the axion backreaction vanishes, while the gauge current approaches the massless limit from above in eq.\ \eqref{gaugebackapprenorm}.  

In  limit $\mu \to \infty$, the physical backreaction currents vanish, $\mathcal{B}_{\rm ren.} \to 0$ and $\mathcal{J}_{\rm ren.}\to 0$, and the equations of motion read
\begin{align}\label{axbackapprenorm}
\bar{Z}_{\phi} (3H\dot{\axion}) +   \frac{dV(\axion)}{d\axion}  + 3(1-2\lambda)\frac{H^4 }{4\pi^2 f} m_Q^3  -\frac{1}{2\pi^2}\frac{\lambda^2 \dot{\phi}}{Hf^2}\left[3-4\left(\frac{\lambda\dot{\phi}}{fH}\right)^2\right]H^4  & =  0,\\
\frac{1}{\bar{g}^2}H^3m_Q \(2 + 2 m_Q^2\) - (1-2\lambda)\frac{H^2}{4\pi^2 f}m_Q^2 \dot{\axion}  &= 0.
\end{align}
Note that the axion gauge field interaction induced in this limit simply shifts the coefficients in the background Chromo-Natural inflation equations. While this effect alters the attractor solution, provided these terms do not cancel,  there is still a Chromo-Natural attractor solution. Backreaction is always unimportant for the gauge field equation of motion, provided the theory remains weakly coupled. In this limit the loops are suppressed by factors of $\alpha = g^2/4\pi \ll 1$. In order that the additional axion contributions do not invalidate the (shifted) Chromo-Natural attractor,  we require 
\begin{align}
\xi = \frac{\lambda\dot{\phi}}{2fH} \ll 1.
\end{align}

In this theoretical infinite-mass limit,  the effects of the massive fermions do not entirely vanish. They leave behind the effective contact interactions in the action in eq.\ \eqref{eqn:lagrangianshift}. In this limit, we reproduce the action for the effective field theory of kinetically driven Chromo-Natural inflation from Ref.\ \cite{Watanabe:2020ctz}. The theory in this limit is described by the one-loop effective action 
\begin{align}\label{eqn:EFTKI}\nn
S_{\rm EFT}= & \int d^4 x \sqrt{-g} \Bigg[\frac{M_{\rm Pl}^2}{2} R -\frac14 F_{\mu\nu}^a  F^{a\mu\nu} +a_1X  + a_2 X^2+a_3 X\Box\phi +a_4 G^{\mu\nu}\partial_\mu\phi\partial_\nu\phi  - V(\axion) \\ & \quad
 +a_5 \phi R_{\rm GB}+a_6 \lambda \phi \epsilon^{\mu\nu\alpha\beta}R^{\rho\sigma}{}_{\mu\nu}R_{\rho\sigma\alpha\beta} + a_7 (\Box\phi)^2 + \frac{\alpha }{4 \pi}\frac{(1-2\lambda) \axion}{ f} F_{\mu\nu}^a  \tilde F^{a\mu\nu}\Bigg]
 \end{align}
 where 
 \begin{align}
 X = \frac{g^{\mu \nu}}{2} \partial_\mu \axion
 \partial_\nu \axion\,,\quad R_{\rm GB}=R^2-4R^{\mu\nu}R_{\mu\nu}+R^{\mu\nu\sigma\rho}R_{\mu\nu\sigma\rho}\,,
 \end{align}
 and for the UV completion above, the parameters are fixed to
\begin{align}
a_1 = 1, \quad a_2 = \frac{2\lambda^4}{3\pi^2 f^4} ,\quad a_3 = 0, \quad  a_4  =  \frac{\lambda^2}{12\pi^2 f^2},\quad  a_5 = 0,\quad  a_6 = \frac{1}{192\pi^2}\frac{\lambda }{f} , \quad a_7 =  - \frac{\lambda^2}{12\pi^2 f^2}.
\end{align}
The authors of Ref.\  \cite{Watanabe:2020ctz} argue that the operator $(\Box\phi)^2$ can be removed using the field equations.

General solutions  (for general $a_i$ in the action in eq.\ \eqref{eqn:EFTKI}) have been explored in detail in Ref. \cite{Watanabe:2020ctz} who find a variety of new attractor solutions. The particular UV completion studied in this work correlates these coefficients, and we leave exploration of the behavior of this realization of the theory in the limit $\xi \gg 1$ for future work. However, we note that from this perspective, it seems difficult to obtain an effective action for Chromo-Natural inflation by integrating out heavy fermions. Recall that Chromo-Natural inflation requires a large $\phi F\tilde{F}$ term, which appears to be difficult to generate from heavy fermions without also inducing these strong derivative self interactions. At finite, but non-zero fermion masses, the result interpolates between the massless and infinite mass limits.

\section{Conclusions}\label{sec:conclusions}

In this work, we have studied the production of both massive and massless fermions in de Sitter space in the presence of a classical non-Abelian SU(2) gauge field and a homogeneous rolling axion. We considered fermions with vector-like gauge field couplings and an axion-dependent mass term. In this form, the axion-fermion interaction is explicitly removed in the  limit where the fermion is massless. The vanishing of the axion effects in the massless limit therefore provides a non-trivial check of our analysis.

To perform computations, it proved most useful to work in the basis where the fermion mass is constant and the axion is coupled derivatively to the axial current of the fermions. This basis is reached by an axion-dependent axial rotation of the fermions. Under such a field transformation, the measure of the path integral is not invariant. We have demonstrated that the additional terms induced in the action are critical to recovering the correct physical behavior in  the massless limit. Conversely, in the very massive limit, the backreaction currents, $\mathcal{B}_{\rm ren.}$ and $\mathcal{J}_{\rm ren.}$ vanish, and the  backreactions is due to the induced contact interactions.

In determining the backreaction currents, we are required to compute divergent quadratic expectation values of quantum fields. To deal with these divergences, we regulate the expressions by adding auxiliary  Pauli-Villars regulator fields whose masses are taken to infinity at the end of the computation. The complicated structure of the equations of motion for the coupled sector unfortunately mean that analytic solutions are not available. However, we developed an extremely accurate approximations to the equations of motion in the UV for both the physical fields, as well as the regulator fields which allowed us to renormalize the theory. We demonstrated that all divergences in the theory can be absorbed into renormalizations of the axion kinetic term, and into the gauge coupling of the SU(2) gauge fields. Our result non-trivially recover the axial anomaly for the divergence of the axial current.

In the limit that the fermions are massless, we found that the interaction of the axion with the fermions cancels completely between the physical currents and the additional terms introduced by the field redefinition. The additional induced gauge field-axion interactions  also cancel between the physical currents and the additional terms induced by the field redefinition. This cancellation provides a very non-trivial check of our results. In this  limit, the massless fermions simply lead to the renormalization of the gauge coupling; perturbation theory remains under control provided we use the value of the gauge coupling evaluated at the scale $m_Q$ when $\mu \ll 1$ or $\mu$ when $\mu > m_Q$.  Provided the gauge coupling does not run to strong coupling at the scale $m_Q$, we find small corrections to the gauge field equations of motion.

In the limit that the fermions are very massive, the contribution of the physical, renormalized currents vanishes leaving behind the finite (fermion) mass-independent contact interaction induced by the field redefinitions. This apparent failure of decoupling is of course familiar from the usual way axion-gauge field interactions are induced. In addition to inducing the usual $\phi F\tilde{F}$ interaction, we find a number of additional axion self interactions. These all involve derivatives of the axion, and thus do not spoil the axion shift symmetry. On the one hand, the appearance of these terms, and their strength, throws into some doubt the self consistency of the original Chromo-Natural inflation model as arising from massive UV fermions.  On the other hand, these additional axion self-interactions may facilitate step slow-roll solutions on their own. We leave investigation of these avenues for future work.

\acknowledgments

We thank Patrick Draper, Aida El Khadra, Marco Peloso, Jessie Shelton, and Lorenzo Sorbo for useful discussions. We additionally thank Lorenzo Sorbo for comments on a draft of this paper. This work was supported in part by the US Department of Energy through grant DE-SC0015655.

\appendix

\setcounter{tocdepth}{-1}

\section{(Anti-)Symmetrization and regularization}\label{app:antisym}

In this appendix we study the anti-symmetrization of quadratic expectation values of operators, as recently advocated in the work of Ref.\ \cite{Mirzagholi:2019jeb}.  We demonstrate that as a consequence of the properties of the solutions under charge-conjugation, that anti-symmetrization does not change the results. 

We begin by considering a Hermitian operator, which is quadratic in the fermionic field, of the general form
\begin{align}\label{eqn:quadfermion}
\mathcal{O}({\bf x}, \eta) = \int {\rm d}^3 k {\rm d}^3 k' e^{i({\bf k} - {\bf k}')\cdot{\bf x}}\psi^\dagger_{{\bf k}} \mathds{A} \psi_{{\bf k'}} = \int {\rm d}^3 k {\rm d}^3 k' e^{i({\bf k} - {\bf k}')\cdot{\bf x}}\psi^\dagger_{{\bf k},\alpha} A_{\alpha, \beta}({\bf k}, {\bf k}', \eta) \psi_{{\bf k'},\beta}
\end{align}
In order to treat particles and anti-particles on equal footing, Ref.\ \cite{Mirzagholi:2019jeb} states that the operator must be anti-symmetrized
\begin{align}\label{eqn:antisym}
\mathcal{O}({\bf x}, \eta) = & \int {\rm d}^3 k {\rm d}^3 k' e^{i({\bf k} - {\bf k}')\cdot{\bf x}}A_{\alpha, \beta} [\psi^\dagger_{{\bf k},\alpha}, \psi_{{\bf k'},\beta}]\\
= & \int {\rm d}^3 k {\rm d}^3 k' e^{i({\bf k} - {\bf k}')\cdot{\bf x}}A_{\alpha, \beta} \frac{1}{2}\(\psi^\dagger_{{\bf k},\alpha} \psi_{{\bf k'},\beta}- \psi_{{\bf k'},\beta}\psi^\dagger_{{\bf k},\alpha}\).
\end{align}
Inserting the mode expansion from eq.\ \eqref{eq:modeexpansion2}, we obtain for the VEV
\begin{align}
\langle\mathcal{O}({\bf x}, \eta) \rangle 
 = & \int {\rm d}^3 k \frac{1}{2}\(V^\dagger_{i} (\eta, - \bt{k} )\mathds{A}V_{i} (\eta, - \bt{k} )-U^\dagger_{i}(\eta,\bt{k} )\mathds{A}U_{i}(\eta,\bt{k} )\)\\
= & \int {\rm d}^3 k  \frac{1}{2}\(U^T_{i} (\eta, \bt{k} )\tilde{\mathds{A}}U^*_{i} (\eta, \bt{k} )-U^\dagger_{i}(\eta,\bt{k} )\mathds{A}U_{i}(\eta,\bt{k} )\),
\end{align}
where
\begin{align}
\tilde{\mathds{A}} = \gamma^0\gamma^5\mathds{A}\gamma^5\gamma^0,
\end{align}
and we have used eq.\ \eqref{eq:rotatedantispinor}. Now, notice that, for the axial charge, $\mathds{A} = \gamma^0\gamma^5$, and therefore
\begin{align}
\label{eq:OAntiSym}
\langle\mathcal{O}({\bf x}, \eta) \rangle
= & -\int {\rm d}^3 k U^\dagger_{i}(\eta,\bt{k} )\gamma^0\gamma^5U_{i}(\eta,\bt{k} ),
\end{align}
which is the same result as would have been obtained by simply evaluating eq.\ \eqref{eqn:quadfermion}. Notice that if we were to evaluate the charge density $j^0 = \bar{\psi} \gamma^0 \psi$, $\mathds{A} = \mathds{1}$, we would find different results using  eq.\ \eqref{eqn:antisym}, compared to evaluating eq.\ \eqref{eqn:quadfermion} directly. In this case the charge density vanishes evaluating the expression \eqref{eqn:antisym}, while is nonzero when evaluated using eq.\ \eqref{eqn:quadfermion}. In reference \cite{Parker:2009uva}, this example is provided as the reasoning behind the antisymmetrization of the current.  

\section{Wentzel-Kramers-Brillouin or adiabatic solutions}\label{app:WKB}

In this appendix, we introduce a Wentzel-Kramers-Brillouin (WKB) expansion of the mode solutions. The expansion parameter is the expansion rate of the Universe, $H = \dot{a}/a$ and we expand the system that follows from the Dirac equations following Refs.\ \cite{Landete:2013lpa, Adshead:2018oaa}. 

\subsection*{Expansion of the solutions}
We seek an adiabatic expansion of the solutions to the equations of motion (eq.\ \eqref{eq:decoupledeoms} for the decoupled sector, and eqs. \eqref{eqn:coupledeoms} and \eqref{eqn:coupledeoms2} for the coupled sector), which schematically take the form
\begin{align}
(i\partial_x +  \mathds{ M}) \vec u =0,
\end{align}
where $\mathds{ M}$ is a matrix that depends on time. We look for an expansion of the solutions in powers of the expansion rate, and thus we rescale the derivative $\partial_x \to T^{-1}\partial_x$ where $T$ is a constant time parameter to characterize the rate of expansion ($T \sim H^{-1}$).

We begin by expanding  $\vec u$ in  powers of $T$ as 
\Beq
\vec u(k,\eta) =\exp\({i\int dx \sum_{m=0} \frac{\omega^{(m)}(x)}{T^{m-1}}  } \) \sum_{n=0} \frac{1}{T^n} \vec u^{(n)}(x).
\Eeq
Substituting into the equations of motion
\Beq
\sum_{n=0}\frac{i \partial_x \vec u^{(n)} }{T^n}- \sum_{m=0} \frac{\omega^{(m)}(x)}{T^{m-1}}  \sum_{n=0} \frac{1}{T^n} \vec u^{(n)}(x) + \sum_{n=0}\frac{1}{T^{n-1}}\mathds{ M}\vec u^{(n)} = 0,
\Eeq
and collecting terms order by order, we obtain
\bea\label{eqn:zerothorder}
&~& (\omega^{(0)}- \mathds{ M} )\vec u^{(0)} =0, \\\label{eqn:orderbyorder}
&~& i \partial_x \vec u^{(k)} -\sum_{m=0}^{m=k+1} {\omega^{(m)}} \vec u^{(k+1-m)} +\mathds{ M}\, u^{(k+1)} =0, \quad k \ge 0. 
\eea
Notice that to solve the zeroth order equation, $\omega^{(0)},\vec u^{(0)}$ are a pair of eigenvalues and eigenvectors of $\mathds{ M}$. For an $N$-dimensional matrix $\mathds{ M}$ we have $N$ such pairs. 

To solve for the first order (in $T^{-1}$) correction, $\omega^{(1)}_i$ and $u^{(1)}_i$, to the $i$th zeroth-order WKB solution, we take $k =0$ in eq.\ \eqref{eqn:orderbyorder} 
\be\label{eqn:1storder}
i \partial_x \vec u^{(0)}_i  -\omega^{(0)}_i \vec u^{(1)}_i- \omega^{(1)}_i \vec u^{(0)}_i+\mathds{ M} \vec u^{(1)}_i=0. 
\ee
Multiplying eq.\ \eqref{eqn:1storder} by $u^{(0)\dag}_i$, and using the Hermitian conjugate of eq.\ \eqref{eqn:zerothorder}, we obtain
\be
\label{eq:WKBu1}
i \vec u^{(0)\dag}_i   \partial_x \vec u^{(0)}_i  -\omega^{(1)}_i \vec u^{(0)\dag}_i \vec u^{(0)}_i=0. \\
\ee
Note that the first term vanishes, hence the first order correction $\omega^{(1)}_i$ to the frequency vanishes. We can then solve for $\vec u^{(1)}_i$
\bea
i \partial_x \vec u^{(0)}_i  -\omega^{(0)}_i\vec u^{(1)}_i +\mathds{ M} \vec u^{(1)}_i=0.
\eea
Since the eigenvectors of $\mathds{ M}$ span a complete $N$-dimensional orthonormal basis, we can write the expansion
\bea
\label{eq:WKBu1exp}
\vec u^{(1)}_i=\sum_{j=1}^N A_{i,j} \vec u^{(0)}_j.
\eea
After plugging this expression back in eq.\ \eqref{eq:WKBu1} and then left-multiplying by $u^{(0)\dag}_{j\neq i}$ we find
\bea
\label{eq:WKBu1A}
A_{i,j\neq i}=\frac{i \vec u^{(0)\dag}_{j}\partial_x \vec u^{(0)}_{i}}{\omega_i^{(0)}-\omega_j^{(0)}}.
\eea
The only coefficient from the expansion in eq.\ \eqref{eq:WKBu1exp} which is undetermined by eq.\ \eqref{eq:WKBu1} is $A_{i,i}$, and we set it to zero without loss of generality. 

Higher order WKB corrections can be computed in a similar manner to the first order ones. For completeness, we give the second order WKB corrections, which we use for the regularization of the current and axion backreaction. We take $k =1$ in eq.\ \eqref{eqn:orderbyorder} 
\bea
\label{eq:WKBu2}
i \partial_x \vec u^{(1)}_i - \omega^{(0)}_i \vec u^{(2)}_i - \omega^{(1)}_i \vec u^{(1)} - \omega^{(2)}_i \vec u^{(0)}_i +\mathds{ M}\, u^{(2)}_i =0.
\eea
After multiplying eq.\ \eqref{eq:WKBu2} by $\vec u^{(0)\dag}_i$ we arrive at (recall $\vec u^{(0)\dag}_i\vec u^{(0)}_j=\delta_{ij}$ and $\omega_i^{(1)}=0$)
\bea
\label{eq:freqWKB2}
\omega_i^{(2)}=i\vec u^{(0)\dag}_i\partial_x \vec u^{(1)}_i.
\eea
To find $\vec u^{(2)}_i$ we again expand in the basis spanned by the eigenvectors of $\mathds{ M}$
\bea
\label{eq:WKBu2exp}
\vec u^{(2)}_i=\sum_{j=1}^N B_{i,j} \vec u^{(0)}_j.
\eea
Upon substituting this expression into eq.\ \eqref{eq:WKBu2} and then left-multiplying by $u^{(0)\dag}_{j\neq i}$ we arrive at
\bea
\label{eq:WKBu2A}
B_{i,j\neq i}=\frac{i \vec u^{(0)\dag}_{j}\partial_x \vec u^{(1)}_{i}}{\omega_i^{(0)}-\omega_j^{(0)}}.
\eea
Again the only coefficient from the expansion in eq.\ \eqref{eq:WKBu2exp} which is unconstrained by eq.\ \eqref{eq:WKBu2} is $B_{i,i}$, and we put it to zero without loss of generality. 

One can continue in this way to compute higher and higher corrections. For the results below, we checked up to fifth order in the expansion. These expressions are easily obtained by repeating the above procedure. We now simply list the results.

Writing,
\begin{align}
\vec u^{(3)} _i = \sum_{k} C_{i,k}\vec u^{(0)} _k, \quad \vec u^{(4)} _i = \sum_{k} D_{i,k}\vec u^{(0)} _k,\quad \vec u^{(5)} _i = \sum_{k} E_{i,k}\vec u^{(0)} _k
\end{align}
we find
\begin{align}
C_{i\neq j} = & ~ \frac{i \vec u^{(0)\dagger}_j \partial_x \vec u_i^{(2)}-{\omega _i ^{(2)}} \vec u^{(0)\dagger}_j \vec u^{(1)}_i}{ ({\omega _i ^{(0)}}-\omega^{(0)}_{j}) }, \quad  C_{i = j} =  0, \quad {\omega _i ^{(3)}}  = i \vec{u}^{(0)\dagger}_i \partial_x \vec u_i^{(2)}-{\omega _i ^{(2)}}\vec{u}^{(0)\dagger}_i   \vec  u^{(1)}_i ,
\end{align}
and 
\begin{align}
D_{i\neq j}   = &~ \frac{i\vec{u}_{j}^{(0)\dagger}\partial_x \vec u_i^{(3)} - \omega^{(2)}_i \vec{u}_{j }^{(0)\dagger} \vec u_i^{(2)}- {\omega_i^{(3)}}\vec{u}_{j}^{(0)\dagger}\vec u_i^{(1)}}{ ({\omega_i^{(0)}} -{\omega_j^{(0)}})}, \quad \omega_i^{(4)}  =  i \vec{u}_i^{(0)\dagger}\partial_x \vec u_i^{(3)} - \omega^{(2)}_i \vec{u}_i^{(0)\dagger} \vec u_i^{(2)}.
\end{align}
Using the normalization condition, and taking the $D_{ij}$ to be real, we set
\begin{align}
D_{i,i}=D^\dagger_{i,i} = - \(\sum_{k}C^{\dagger}_{i,k}A_{i,k}+\sum_{k}A^{\dagger}_{i,k}C_{i,k} +\sum_{k}B^{\dagger}_{i,k}B_{i,k}\).
\end{align}
Finally, 
\begin{align}
 {\omega_i^{(5)}}   = & i \vec{u}^{(0)\dagger}_i\partial_x \vec u_i^{(4)}   - {\omega_i^{(3)}} \vec{u}^{(0)\dagger}_i \vec u_i^{(2)},
\end{align}
and 
\begin{align}
E_{i\neq j}  = &~ \frac{i \vec{u}^{(0)\dagger}_j\partial_x \vec u_i^{(4)} - {\omega_i^{(4)}}\vec{u}^{(0)\dagger}_j \vec u_i^{(1)} - {\omega_i^{(3)}} \vec{u}^{(0)\dagger}_j \vec u_i^{(2)} - {\omega_i^{(2)}} \vec{u}^{(0)\dagger}_j \vec u_i^{(3)} }{({\omega_i^{(0)}}-{\omega_j^{(0)}}) }, \quad E_{ii} = 0.
\end{align}

\subsection*{Lowest order solutions: Decoupled sector I, $\{u_1^+, v_1^{+}\}$}\label{sec:Dec1}
The first decoupled sector consists of two modes 
\bea
\vec u = \frac{1}{\sqrt{2}}\(\begin{matrix}u_1^+ \\ v_1^+ \end{matrix}\),
\eea
that are governed by the two-dimensional matrix
\Beq
\mathds{ M}=-\sigma^3+\frac{1}{x}\left[\left(\frac{m_Q}{2}-2\xi\right)\sigma^3+\mu\sigma^1\right].
\Eeq
Its eigenvalues are
\Beq
\omega_1^{(0)}&=\frac{1}{2 x}\sqrt{4 \mu ^2+\left(m_Q-2 (2 \xi +x)\right){}^2},\\
\omega_2^{(0)}&=-\frac{1}{2 x}\sqrt{4 \mu ^2+(m_Q-2 (2 \xi +x))^2}.
\Eeq
There are analytic expressions for the eigenvectors $\vec u^{(0)}_1$ and $\vec u^{(0)}_2$, which we do not provide here (and similarly for the other sectors in the sections below) in order to avoid clutter. 

We take $\omega^{(0)}_1$ and its corresponding eigenvector, $\vec u^{(0)}_1$, as the zeroth-order positive-frequency WKB solution. The first- and second-order corrections to the positive-frequency WKB solution are found according to eq.\ \eqref{eq:WKBu1A} and eqs.\ \eqref{eq:freqWKB2} and \eqref{eq:WKBu2A}, respectively.

\subsection*{Lowest order solutions: Decoupled sector II $\{u_2^-, v_2^{-}\}$}\label{sec:Dec2}
The other decoupled modes
\bea
\vec u = \frac{1}{\sqrt{2}}\(\begin{matrix}u_2^- \\ -v_2^- \end{matrix}\),
\eea
are governed by the two-dimensional matrix
\Beq
\mathds{ M}=\sigma^3+\frac{1}{x}\left[\left(\frac{m_Q}{2}-2\xi\right)\sigma^3+\mu\sigma^1\right].
\Eeq
Its eigenvalues are
\Beq
\omega_1^{(0)}&=\frac{1}{2 x}\sqrt{4 \mu ^2+\left(m_Q-2( 2\xi - x)\right){}^2},\\
\omega_2^{(0)}&=-\frac{1}{2 x}\sqrt{4 \mu ^2+\left(m_Q-2( \xi - x)\right){}^2}.
\Eeq
Again $\omega^{(0)}_1$ and its corresponding eigenvector, $\vec u^{(0)}_1$, give the zeroth-order positive-frequency WKB solution, and can be used to find the first- and second-order WKB corrections according to eq.\ \eqref{eq:WKBu1A} and eqs.\ (\ref{eq:freqWKB2},\ref{eq:WKBu2A}), respectively.

\subsection*{Lowest order solutions: Coupled sector $\{u_1^-, v_1^{-}\}$ and $\{u_2^{+}, v_{2}^{+}\}$}\label{sec:CoupledSector}
The coupled modes
\bea
\vec u = \frac{1}{\sqrt{2}}\(\begin{matrix}u_1^- \\ -v_1^- \\u_2^+ \\ v_2^+ \end{matrix}\),
\eea
are governed by the four-dimensional matrix
\Beq\label{eqn:Mcoupled}
\mathds{ M}=\left(
\begin{array}{cccc}
 -\dfrac{m_Q}{2 x}-\dfrac{2 \xi }{x}+1 & \dfrac{\mu }{x} & \dfrac{m_Q}{x} & 0 \\
 \dfrac{\mu }{x} & \dfrac{m_Q}{2 x}+\dfrac{2 \xi }{x}-1 & 0 & -\dfrac{m_Q}{x} \\
 \dfrac{m_Q}{x} & 0 & -\dfrac{m_Q}{2 x}-\dfrac{2 \xi }{x}-1 & \dfrac{\mu }{x} \\
 0 & -\dfrac{m_Q}{x} & \dfrac{\mu }{x} & \dfrac{m_Q}{2 x}+\dfrac{2 \xi }{x}+1 \\
\end{array}
\right).
\Eeq
Its eigenvalues are
\begin{align}\nn
\omega_1^{(0)} = & \frac{\sqrt{\mu^2+\(\sqrt{x^2+m_Q^2}+\(\frac{m_Q}{2}+2\xi\)\)^2}}{x} ,\quad  \omega_3^{(0)} =  - \frac{\sqrt{\mu^2+\(\sqrt{x^2+m_Q^2}+\(\frac{m_Q}{2}+2\xi\)\)^2}}{x}, \\
\omega_2^{(0)} = & \frac{\sqrt{\mu^2+\(\sqrt{x^2+m_Q^2}-\(\frac{m_Q}{2}+2\xi\)\)^2}}{x}
 , \quad
\omega_4^{(0)} = - \frac{\sqrt{\mu^2+\(\sqrt{x^2+m_Q^2}-\(\frac{m_Q}{2}+2\xi\)\)^2}}{x}. 
\end{align}

We take $\omega^{(0)}_1$ and $\omega^{(0)}_2$, and their corresponding eigenvectors, $\vec u^{(0)}_1$, and , $\vec u^{(0)}_2$, respectively, as the two zeroth-order positive-frequency WKB solutions. Again, the first- and second-order WKB corrections are found from eq.\ \eqref{eq:WKBu1A} and eqs.\ \eqref{eq:freqWKB2}) and \eqref{eq:WKBu2A}, respectively.

\section{Series solutions to the coupled sector}\label{app:seriessols}

We consider the coupled equations of motion in eq.\ \eqref{eqn:coupledeoms}. Working with $x = -k\tau$, and writing $M = 2 \xi + m_Q/2$, and
 rescaling
\begin{align}
\(\begin{matrix}U^{-}_1 \\ V^{-}_1\\ U_{2+} \\ V_{2+}\end{matrix}\)  = \sqrt{x}\(\begin{matrix}u^{-}_1 \\ -v^{-}_1\\u_{2+} \\ v_{2+}\end{matrix}\) ,
\end{align}
then switching variable to $u = 2 i x$, the equations can be written
\begin{align}
\(\mathds{1}\partial_u + \mathbb{M}\)\(\begin{matrix}U^{-}_1 \\ V^{-}_1\\ U_{2+} \\ V_{2+}\end{matrix}\)  = 0,
\end{align}
where
\begin{align}
 \mathbb{M}= \(\begin{matrix}  -\frac{1}{2u}- \(\frac{1}{2} -i \frac{M}{u}\) & - i\frac{\mu}{u} & -i\frac{m_Q}{u} & 0  \\
 -i\frac{\mu}{u}  &-\frac{1}{2u}+\(\frac{1}{2} -i \frac{M}{u}\) & 0 &  i\frac{m_Q}{u}  \\  -i\frac{m_Q}{u}  & 0 &    -\frac{1}{2u} + \(\frac{1}{2} +i \frac{M}{u}\) & - i\frac{\mu}{u} \\ 
0 &  i\frac{m_Q}{u}  &-i\frac{\mu}{u} &  -\frac{1}{2u} - \(\frac{1}{2} +i \frac{M}{u}\) 
 \end{matrix}\) .
\end{align}
These equations can be partially decoupled, by operating with the conjugate of the differential operator to obtain
\begin{align}\label{eqn:seconddecouped}
U^-_1{}''+\frac{\kappa-1   }{u}U^-_1+  \[-\frac{1}{4}+\frac{\frac{1}{4}-\tilde{\mu}^2}{u^2}\]U^-_1 = \frac{2 M m_Q}{u^2} U^+_2,\\
U^+_2{}''+\frac{1-\kappa}{u}U^+_2+  \[-\frac{1}{4}+\frac{\frac{1}{4}-\tilde{\mu}^2}{u^2}\]U^+_2 = \frac{2 M  m_Q}{u^2} U^-_1,\\
   V^-_1{}''+\frac{\kappa
   }{u}V^-_1+  \[-\frac{1}{4}+\frac{\frac{1}{4}-\tilde{\mu}^2}{u^2}\]V^-_1 = \frac{2 M  m_Q}{u^2} V^+_2,\\
V^+_2{}''-\frac{\kappa}{u}V^+_2+  \[-\frac{1}{4}+\frac{\frac{1}{4}-\tilde{\mu}^2}{u^2}\]V^+_2 = \frac{2 M  m_Q}{u^2} V^-_1,
\end{align}
where
 \begin{align}
\kappa = \frac{1}{2}+i M,\quad  \tilde{\mu}^2 = -\(\mu^2+ m_Q^2+M^2\), \quad M = 2 \xi + \frac{m_Q}{2}.
\end{align}

\subsection*{Solutions at early times}

The point $u = \infty$ is an irregular singular point of the equations in eq.\ \eqref{eqn:seconddecouped}, and as such we can look for a (possibly asymptotic) series solution about this point.  We start with the equations for $V$, and look for series solution of the form
\begin{align}
\[\begin{matrix}V^-_1\\V^+_2 \end{matrix} \] = e^{\gamma u}u^\lambda \sum_{n = 0}^{\infty} \[\begin{matrix}a_n\\b_n \end{matrix} \]u^{-n},
\end{align}
where we take $u$ to be large. Substituting into  eq.\ \eqref{eqn:seconddecouped}, we obtain
\begin{align}\nn
\sum_{n = 0}^{\infty} \[\gamma ^2   u^{2 }+(\lambda -n-1) (\lambda -n)  +2 \gamma  (\lambda -n)   u+ \kappa u-\frac{1}{4}u^{2}+\(\frac{1}{4}-\tilde{\mu}^2\)\]\frac{a_n}{u^{n}} = & \sum_{n = 0}^{\infty}2 M m_Q \frac{b_n}{u^{n}} ,\\
\sum_{n = 0}^{\infty} \[\gamma ^2  u^{2 }+(\lambda -n-1) (\lambda -n) +2 \gamma  (\lambda -n) u- \kappa u-\frac{1}{4} u^{2}+\(\frac{1}{4}-\tilde{\mu}^2\) \] \frac{b_n}{u^{n}}  = & \sum_{n = 0}^{\infty}2 M m_Q \frac{a_n}{u^{n}}.
\end{align}
In order that the most divergent terms satisfy the equations, those proportional to $u^2$, we require
\begin{align}
\gamma = \pm \frac{1}{2}.
\end{align}
The terms proportional to $u$ imply 
\begin{align}
(2 \gamma  \lambda  + \kappa )a_0  = 0,\\
(2 \gamma  \lambda - \kappa )b_0  = 0.
\end{align}
We therefore have two sets of solutions
\begin{align}
\lambda = -\frac{\kappa}{2\gamma} = \mp \kappa, \quad b_0 = 0, \quad a_0 \neq 0,\\
\lambda = \frac{\kappa}{2\gamma} = \pm \kappa, \quad b_0 \neq 0, \quad a_0 = 0.
\end{align}
We can then find the recursion for the $a_n$ and $b_n$, which reads 
\begin{align}
\[\begin{matrix} a_{n+1}\\b_{n+1}\end{matrix}\] = \mathbb{B}_{n}\[\begin{matrix} a_{n}\\b_{n}\end{matrix}\],
\end{align}
where
\begin{align}
\mathbb{B}_{n} = &- \[\begin{matrix}\frac{(\lambda -n-1) (\lambda -n) +\(\frac{1}{4}-\tilde{\mu}^2\)}{\(2 \gamma  (\lambda -n-1)  + \kappa \)}  &-\frac{2 M m_Q}{\(2 \gamma  (\lambda -n-1)  + \kappa \)}  \\-\frac{2 M m_Q}{\(2 \gamma  (\lambda -n-1)  - \kappa \)}  & \frac{(\lambda -n-1) (\lambda -n) +\(\frac{1}{4}-\tilde{\mu}^2\)}{\(2 \gamma  (\lambda -n-1)  - \kappa \)} \end{matrix}\].
\end{align}

The solutions for $U_1^-$ and $U_2^+$ which we write
\begin{align}
\[\begin{matrix}U^-_1\\U^+_2 \end{matrix} \] = e^{\gamma u}u^{\sigma} \sum_{n = 0}^{\infty} \[\begin{matrix}c_n\\d_n \end{matrix} \]u^{-n},
\end{align}
can be found by replacing $\kappa \to \kappa - 1$, and replacing $\lambda \to \sigma$ above. Again, we have
\begin{align}
\gamma = \pm \frac{1}{2}.
\end{align}
The next most divergent terms read
\begin{align}
(2 \gamma  \sigma  + (\kappa -1))c_0  = 0,\\
(2 \gamma   \sigma - (\kappa-1) )d_0  = 0.
\end{align}
Note that both $c_0\neq 0$ and $d_0 \neq 0$. So we have the sets of solutions
\begin{align}
\sigma = -\frac{\kappa-1}{2\gamma} = \mp (\kappa-1), \quad d_0 = 0, \quad c_0 \neq 0,\\
\sigma = \frac{\kappa-1}{2\gamma} = \pm (\kappa-1), \quad c_0 \neq 0, \quad d_0 = 0.
\end{align}
The series is then generated by the recursion
\begin{align}
\[\begin{matrix} c_{n+1}\\d_{n+1}\end{matrix}\] = \mathbb{C}_{n}\[\begin{matrix} c_{n}\\d_{n}\end{matrix}\].
\end{align}
where
\begin{align}
\mathbb{C}_{n} 
= &- \[\begin{matrix}\frac{(\sigma -n-1) (\sigma -n) +\(\frac{1}{4}-\tilde{\mu}^2\)}{\(2 \gamma  (\sigma -n-1)  + \kappa-1 \)}  &-\frac{2 M m_Q}{\(2 \gamma  (\sigma -n-1)  + \kappa -1\)}  \\-\frac{2 M m_Q}{\(2 \gamma  (\sigma -n-1) +1 - \kappa \)}  & \frac{(\sigma -n-1) (\sigma -n) +\(\frac{1}{4}-\tilde{\mu}^2\)}{\(2 \gamma  (\sigma -n-1)  +1- \kappa \)} \end{matrix}\].
\end{align}

To find the final solutions, we need to impose the 1st-order equation. These must have the same asymptotic divergence, but may have different power-law divergence. We write
\begin{align}
\[\begin{matrix} U_1^- \\ V_1^- \\ U_2^+ \\ V_2^+ \end{matrix}\] = e^{\gamma u}\sum_{n = 0}^{\infty}\[\begin{matrix}c_n x^{\sigma} \\ a_n x^{\lambda} \\ d_n x^{\sigma}\\ b_n x^{\lambda} \end{matrix}\]x^{-n},
\end{align}
plugging in 
\begin{align}
\sum_{n = 0}^{\infty}\left(
\begin{array}{c}
 u^{\sigma } \left(-c_n (2 n-2 \sigma -2iM - 2 \gamma  u+u+1)- 2i  m_Q d_n\right)-2 i \mu a_n u^{\lambda } \\
u^{\lambda } \left(a_n \left(2 \lambda -2i M -2 n+2 \gamma  u+u-1\right)+2 i b_n m_Q\right)-2 i \mu c_n u^{\sigma } \\
u^{\sigma } \left(d_n (-2 n+2 \sigma +2iM+2 \gamma  u+u-1)-2i m_Q c_n\right)-2 i \mu b_n u^{\lambda }\\
u^{\lambda } \left(2 i a_n m_Q-b_n \left(-2 \lambda +2i M+2 n-2 \gamma  u+u+1\right)\right)-2 i \mu d_n u^{\sigma } \\
\end{array}
\right) u^{-n} =0.
\end{align}
We want solutions that match onto the Bunch-Davies vacuum (modes with frequency $e^{ix}$), so we take $\gamma = 1/2$. From above, setting $\gamma = 1/2$, we find the vectors that generate the asymptotic series, and changing variable back to $x = u/2i$, we find the solutions
\begin{align}
\psi_1 & =  \sqrt{2}e^{ix}  x^{-i M}\sum_{n = 0}^{\infty}(2 i x)^{-n}\(\begin{matrix}c^{(1)}_n  \\ a^{(1)}_n \\ d^{(1)}_n \\ b^{(1)}_n  \end{matrix}\) , \quad \(\begin{matrix}c^{(1)}_0  \\ a^{(1)}_0 \\ d^{(1)}_0 \\ b^{(1)}_0  \end{matrix}\)  = \(\begin{matrix} 1 \\ \frac{i\mu}{2 i x} \\ 0 \\ 0 \end{matrix} \),\\
\psi_2 & =  \sqrt{2}e^{ix}  x^{iM}\sum_{n = 0}^{\infty}(2 i x)^{-n}\(\begin{matrix}c^{(2)}_n  \\ a^{(2)}_n \\ d^{(2)}_n \\ b^{(2)}_n  \end{matrix}\), \quad \(\begin{matrix}c^{(2)}_0  \\ a^{(2)}_0 \\ d^{(2)}_0 \\ b^{(2)}_0  \end{matrix}\)  =\(\begin{matrix}0 \\ 0 \\  \frac{i\mu}{2 i x}\\ 1\end{matrix} \) .
\end{align}
The coefficients are generated by the recursion relations
\begin{align}
\(\begin{matrix}c^{(i)}_{n+1}  \\ a^{(i)}_{n+1} \\ d^{(i)}_{n+1} \\ b^{(i)}_{n+1}  \end{matrix}\) = \mathbb{D}^{(i)}_{n}\(\begin{matrix}c^{(i)}_{n}  \\ a^{(i)}_n \\ d^{(i)}_n \\ b^{(i)}_n  \end{matrix}\),
\end{align}
where $i \in\{1, 2\}$ and the matrix $\mathbb{D}^{(i)}_{n}$ is
\begin{align}
\mathbb{D}^{(i)}_{n}  = \[\begin{matrix}
\frac{(\sigma -n-1) (\sigma -n) +\(\frac{1}{4}-\tilde{\mu}^2\)}{\(2 \gamma  (1+n-\sigma )  - \kappa+1 \)}  & 0 & -\frac{2 M m_Q}{\(2 \gamma  (1+n-\sigma )  - \kappa+1\)}  & 0 \\
0 & \frac{(\lambda -n-1) (\lambda -n) +\(\frac{1}{4}-\tilde{\mu}^2\)}{\(2 \gamma  (1+n-\lambda)  - \kappa \)}  & 0 & -\frac{2 M m_Q}{\(2 \gamma  (1+n-\lambda)  - \kappa \)}  \\ 
-\frac{2 M m_Q}{\(2 \gamma  ( 1+n-\sigma) -1 + \kappa\)} & 0 & \frac{(\sigma -n-1) (\sigma -n) +\(\frac{1}{4}-\tilde{\mu}^2\)}{\(2 \gamma  ( 1+n-\sigma) -1 + \kappa \)} & 0 \\ 
0 & -\frac{2 M m_Q}{\(2 \gamma  (1+n-\lambda)  + \kappa \)}  & 0 &  \frac{(\lambda -n-1) (\lambda -n) +\(\frac{1}{4}-\tilde{\mu}^2\)}{\(2 \gamma  (1+n-\lambda)  + \kappa  \)} \end{matrix}\]
\end{align}
where $\lambda$ and $\sigma$ take values
\begin{align}
i = 1;\quad  \lambda = -\kappa,\quad  \sigma = \kappa-1,\\
i = 2; \quad \lambda = \kappa,\quad  \sigma = 1-\kappa.
\end{align}
Solutions can also be found in this way for late times, $x =  -k\tau \to 0$. We do not use these solutions in this work, and omit them here.

\section{Solutions to the coupled sector in the large fermion mass limit}\label{app:PVsols}

In this appendix, we expand the adiabatic or WKB solutions from appendix \ref{app:WKB} in the large mass limit in order to find analytic expressions for the contributions of the regulator fields to the currents in the main text. 
 
In order to use Paul-Villars regularization, we require accurate solutions to the equations of motion in the large mass limit in the region where $x < \mu$ as well as where $\mu < x$. Unfortunately, the series solutions found above in appendix \ref{app:seriessols} are accurate only in the second regime. As pointed out in Ref.\ \cite{Weinberg:2010wq}, adiabatic, or WKB solutions can be used to approximate the solutions to the equations of motion for the heavy regulator fields necessary for Pauli-Villars regularization. Unfortunately, the naive application of the results from appendix  \ref{app:WKB} quickly results in lengthy, messy expressions due to the complicated basis eigenvectors $\vec{u}^{(0)}_i$.   

The WKB expansion is effectively an expansion in inverse powers of the frequency, and thus in the large mass limit, this becomes an expansion in inverse powers of the mass.  We can accurately approximate the contribution from the regulator fields in the large mass limit using a series expansion of the WKB solutions in powers of the inverse mass.  However, we need to remain agnostic as to whether $x = -k\tau$ is larger or smaller than $\mu$. We proceed by rescaling the system and working with the variable $y = x/\mu$, and expand the solutions in a series in $\varepsilon = 1/\mu$, while holding $y$ fixed.  

\subsection*{Basis eigenvectors}

In terms of $y$ and $\varepsilon$, the eigenvalues for the matrix $\mathds{M}$ above in eqn \ref{eqn:Mcoupled} are then
\begin{align}
\omega_1^{(0)} = \pm\frac{\sqrt{1+\(\sqrt{y^2+m_Q^2\varepsilon^2}-\gamma \varepsilon\)^2}}{y}, \quad
\omega_2^{(0)} = \pm\frac{\sqrt{1+\(\sqrt{y^2+m_Q^2\varepsilon^2}+\gamma\varepsilon\)^2}}{y},
\end{align}
where 
\begin{align}
\gamma = \frac{m_Q}{2}+2\xi.
\end{align}
We expand in the limit $\varepsilon \ll 1$, to obtain
\begin{align}
\omega_1 = & \frac{\sqrt{y^2+1}}{y}-\frac{\gamma   }{\sqrt{y^2+1}}\varepsilon+\frac{
   \left(\gamma ^2+m_Q ^2+m_Q ^2 y^2\right)}{2 y
   \left(y^2+1\right)^{3/2}}\varepsilon ^2+O\left(\varepsilon ^3\right) ,\\
\omega_2 =& \frac{\sqrt{y^2+1}}{y}+\frac{\gamma   }{\sqrt{y^2+1}}\varepsilon +\frac{
   \left(\gamma ^2+m_Q ^2+m_Q ^2 y^2\right)}{2 y
   \left(y^2+1\right)^{3/2}}\varepsilon ^2+O\left(\varepsilon ^3\right) .
\end{align}
We can similarly find an expansion for the corresponding eigenvectors. Writing $\omega  = \sqrt{y^2+1}/y$, we have at zeroth order
\begin{align}\label{eqn:basis1}
\tilde{\vec{u}}_1 = \frac{1}{\sqrt{2}}\left(
\begin{array}{c}
 -\frac{1}{\sqrt{ 1+y^2 \left(\omega+1\right)}} \\
 \frac{1}{ \sqrt{1-y^2(\omega-1)}} \\
 0 \\
 0 \\
\end{array}
\right) ,\quad 
\tilde{\vec{u}}_2 =\frac{1}{\sqrt{2}}\left(
\begin{array}{c}
 \frac{y(\omega+1)}{\sqrt{y^2(\omega+1)+1}} \\
 \frac{1}{\sqrt{ y^2(\omega+1)+1}} \\
 0 \\
 0 \\
\end{array}
\right),
\end{align}
and
\begin{align}\label{eqn:basis2}
\tilde{\vec{u}}_3 = &\frac{1}{\sqrt{2}} \left(
\begin{array}{c}
 0 \\
 0 \\
 \frac{-y(\omega+1)}{\sqrt{ y^2(\omega+1)+1}} \\
 \frac{1}{\sqrt{y^2(\omega+1)+1}} \\
\end{array}
\right),\quad \tilde{\vec{u}}_4 =\frac{1}{\sqrt{2}} \left(
\begin{array}{c}
 0 \\
 0 \\
 \frac{1}{ \sqrt{y^2\omega^2\left(\frac{1}{\sqrt{\omega}}+1\right)}} \\
\sqrt{\frac{1}{\omega}+1}\\
\end{array}
\right).
\end{align}
Note that
\begin{align}\label{eqn:derivrel}
\partial_y \vec{u}_2 = \frac{1}{2} \frac{1}{y\omega^2} \vec{u}_1, \quad 
\partial_y \vec{u}_1 =   -\frac{1}{2}\frac{1}{y\omega^2}\vec{u}_2, \quad
\partial_y \vec{u}_3 =  \frac{1}{2}\frac{1}{y\omega^2}  \vec{u}_4, \quad
\partial_y \vec{u}_4 =  -\frac{1}{2}\frac{1}{y\omega^2} \vec{u}_3.
\end{align}
The vectors $\vec{u}_i$ form a complete orthonormal basis, and it proves useful for what follows to use this basis to expand the full eigenvectors of $\mathds{ M}$ in a series in powers of $\varepsilon$. We require the solutions up to order $\varepsilon^4$. We write
\begin{align}
\vec{u}_i =\tilde{\vec{u}}_i +  \sum_k(\varepsilon W_{i,k} \tilde{\vec{u}}_k+  \varepsilon^2 X_{i,k} \tilde{\vec{u}}_k + \varepsilon^3 Y_{i,k} \tilde{\vec{u}}_k +  Z_{i,k} \tilde{\vec{u}}_k  \varepsilon^4)+\mathcal{O}(\varepsilon^5) ,
\end{align}
where the $\tilde{u}_j$ are the basis vectors above. The coefficients functions can be found by taking inner products of the zeroth order $\tilde{\vec{u}}_i$ with the full eigenvectors $\vec{u}_j$. Note that the coefficient functions obey consistency relations imposed by orthnormalization conditions
\begin{align}
\vec{u}^\dagger_i \cdot \vec{u}_j = \delta_{ij} .
\end{align}
We list the lowest order terms here,
\begin{align}
[W] = \left(
\begin{array}{cccc}
 0 & -\frac{\gamma }{2 y^2+2} & \frac{m_Q }{2 y \sqrt{y^2+1}} & \frac{m_Q }{2 \sqrt{y^2+1}} \\
 \frac{\gamma }{2 y^2+2} & 0 & -\frac{m_Q }{2 \sqrt{y^2+1}} & \frac{m_Q }{2 y \sqrt{y^2+1}} \\
 -\frac{m_Q }{2 y \sqrt{y^2+1}} & \frac{m_Q }{2 \sqrt{y^2+1}} & 0 & -\frac{\gamma }{2 \left(y^2+1\right)} \\
 -\frac{m_Q }{2 \sqrt{y^2+1}} & -\frac{m_Q }{2 y \sqrt{y^2+1}} & \frac{\gamma }{2 y^2+2} & 0 \\
\end{array}
\right),
\end{align}
and
\begin{align}
[X] = \left(
\begin{array}{cccc}
 -\frac{\gamma ^2}{8 \left(y^2+1\right)^2}-\frac{m_Q ^2}{8 y^2} & \frac{m_Q ^2+y^2 \left(m_Q ^2-2 \gamma ^2\right)}{4 y
   \left(y^2+1\right)^2} & \frac{\gamma  m_Q }{4 \left(y^2+1\right)^{3/2}} & -\frac{\gamma  m_Q }{4 y \left(y^2+1\right)^{3/2}} \\
 \frac{2 \gamma ^2 y^2-m_Q ^2 \left(y^2+1\right)}{4 y \left(y^2+1\right)^2} & -\frac{\gamma ^2}{8 \left(y^2+1\right)^2}-\frac{m_Q ^2}{8 y^2} &
   \frac{\gamma  m_Q }{4 y \left(y^2+1\right)^{3/2}} & \frac{\gamma  m_Q }{4 \left(y^2+1\right)^{3/2}} \\
 \frac{\gamma  m_Q }{4 \left(y^2+1\right)^{3/2}} & \frac{\gamma  m_Q }{4 y \left(y^2+1\right)^{3/2}} & -\frac{\gamma ^2}{8
   \left(y^2+1\right)^2}-\frac{m_Q ^2}{8 y^2} & \frac{2 \gamma ^2 y^2-m_Q ^2 \left(y^2+1\right)}{4 y \left(y^2+1\right)^2} \\
 -\frac{\gamma  m_Q }{4 y \left(y^2+1\right)^{3/2}} & \frac{\gamma  m_Q }{4 \left(y^2+1\right)^{3/2}} & \frac{m_Q ^2+y^2 \left(m_Q ^2-2
   \gamma ^2\right)}{4 y \left(y^2+1\right)^2} & -\frac{\gamma ^2}{8 \left(y^2+1\right)^2}-\frac{m_Q ^2}{8 y^2} \\
\end{array}
\right).
\end{align}
The higher order terms are straightforward albeit messy to compute, and we omit them.

\subsection*{WKB corrections}

We can now use the basis above in eq.\ \eqref{eqn:basis1} and \eqref{eqn:basis2} to expand the WKB correction above in eqs.\ \eqref{eq:WKBu1A}\footnote{In this section, we do not use the Einstein summation convention---repeated indices are not summed unless noted otherwise.}
\begin{align}
A_{i,j\neq i}= \varepsilon \frac{i \vec u^{(0)\dag}_{j}\partial_y \vec u^{(0)}_{i} }{\omega_i^{(0)}-\omega_j^{(0)}}  = i\varepsilon \frac{U_{ji}}{\omega_i^{(0)}-\omega_j^{(0)}} .
\end{align}
We write
\begin{align}
D_{ij} = \tilde{\vec u}^{(0)\dag}_{i} \partial_y \tilde{\vec u}^{(0)}_{j} 
\end{align}
which, due to the relations in eq.\ \eqref{eqn:derivrel} above is of the simple form
\begin{align}
[D]= \left(
\begin{array}{cccc}
 0 & -\frac{1}{2 y^2+2} & 0 & 0 \\
 \frac{1}{2 y^2+2} & 0 & 0 & 0 \\
 0 & 0 & 0 & \frac{1}{2 y^2+2} \\
 0 & 0 & -\frac{1}{2 y^2+2} & 0 \\
\end{array}
\right).
\end{align}
Then,
\begin{align}\nn
U_{ji}   = &  ( D_{ji}+ \varepsilon (   D_{jm}W_{i,m}+   W^\dagger_{j,k} D_{ki})+\varepsilon^2(  D_{jm}X_{i,m} +W^\dagger_{j,k}  D_{km}  W_{i,m}+   X^\dagger_{j,k}D_{ki})\\\nn
  &+ \varepsilon^3(   Y_{i,m}D_{jm} +X^\dagger_{j,k} D_{km}  W_{i,m}+W^\dagger_{j,k}  D_{km} X_{i,m}+   Y^\dagger_{j,k} D_{ki} )\\\nn
   &+ \varepsilon^4(   Z_{i,m}D_{jm} + Y^\dagger_{j,k} D_{km}  W_{i,m}+W^\dagger_{j,k}  D_{km} Y_{i,m}+   X^\dagger_{j,k}  D_{km} X_{i,m}+Z^\dagger_{j,k} D_{ki} )\\\nn
  & +  \varepsilon \partial_y  W_{i,j}+\varepsilon^2 ( \partial_y X_{i,j} +  W^\dagger_{j,m}  \partial_y  W_{i,m}) + \varepsilon^3 ( \partial_y Y_{i,j} +  X^\dagger_{j,m}  \partial_y  W_{i,m}+W^\dagger_{j,m}  \partial_y  X_{i,m})\\
 &  + \varepsilon^4 ( \partial_y Z_{i,j} +  Y^\dagger_{j,m}  \partial_y  W_{i,m}+W^\dagger_{j,m}  \partial_y  Y_{i,m} +X^\dagger_{j,m}  \partial_y  X_{i,m} )  + \mathcal{O}(\varepsilon^5),
  \end{align}
  where we have suppressed sum notation to avoid clutter; the indices $m$ and $k$ are summed, but $i$ and $j$ are not. 
  
The higher order corrections then read
\bea
\omega_i^{(2)} = i \varepsilon \sum_k A_{i,k}  U_{ik},
\eea
where we used the fact that $A_{i,i} = 0$. The off-diagonal mode-corrections are
\begin{align}
B_{i,j\neq i}  
= & \frac{i\varepsilon } {\omega_i^{(0)}-\omega_j^{(0)}} \(  \sum_kA_{i,k} U_{jk} +\partial_y A_{i,j}\).
\end{align}
The third order correction to the frequency reads
\begin{align}
&{\omega _i ^{(3)}}  = i\epsilon\partial_y B_{i,i}.
\end{align}
Substituting in the expansions above, we have
\begin{align}
C_{i\neq j} = & ~i  \frac{ \varepsilon (\sum_{k} B_{i,k} U_{jk} + \partial_y B_{i,j} )+ i{\omega _i ^{(2)}}A_{i,j}}{ ({\omega _i ^{(0)}}-\omega^{(0)}_{j}) } ,
\end{align}
and recall we set $C_{i,i} = 0$. The fourth order correction to the frequency reads
\begin{align}
\omega_i^{(4)}= & ~ i \varepsilon  \sum_k C_{i,k}\vec{u}_i^{(0)\dagger}\partial_y \vec u_k^{(0)} - \omega^{(2)}_i B_{i,i}.
\end{align}
The off-diagonal mode corrections read
\begin{align}
D_{i\neq j} = & ~i  \frac{ \varepsilon ( \sum_{k} C_{i,k} U_{jk} + \partial_y C_{i,j} )+ i{\omega _i ^{(2)}}B_{i,j}}{ ({\omega _i ^{(0)}}-\omega^{(0)}_{j}) } ,
\end{align}
where we made use of the fact that $\omega^{(3)}_i = 0$. The diagonal terms are
\begin{align}
D_{i,i}=D^\dagger_{i,i} = - \(\sum_{k}C^{\dagger}_{i,k}A_{i,k}+\sum_{k}A^{\dagger}_{i,k}C_{i,k} +\sum_{k}B^{\dagger}_{i,k}B_{i,k}\).
\end{align}

\subsection*{The currents}

Now that we have the expansions, it is straightforward to compute the contributions of the regulator fields to the currents. The gauge current is
\begin{align} 
\mathcal{J}_n =  -\frac{\delta_i^a}{3a^3} \langle \bar{\psi}_n \gamma^i \tau^a \psi_n \rangle =  \frac{\delta_i^a}{6a^3} \int \frac{d^3 k}{(2\pi)^3} \sum_j \bar{V}_{j , n}\gamma^i \tau^a V_{j, n} = \frac{\delta_i^a H^3 }{3}\frac{1}{2\pi^2} \varepsilon^{-3} \int y^3 {\rm d}\ln y {\rm J}_n(y),
\end{align}
where (using \textsc{Mathematica}) we find
\begin{align}\nn
 {\rm J}_{n}(x) = &Z_n^{-1}\Bigg(\varepsilon  \left(\frac{2 \left(\gamma +2 \left(y^2+1\right) m_Q \right)}{\left(y^2+1\right)^{3/2}}\right)\\ \nn & +\varepsilon ^3 \left(-\frac{\left(1-4 y^2\right) \gamma ^3+6 \left(y^2+1\right) m_Q  \gamma ^2+3 \left(y^2+1\right) m_Q
   ^2 \gamma +2 \left(y^2+1\right)^2 m_Q ^3}{\left(y^2+1\right)^{7/2}}\)\\ & +\varepsilon^3\(-\frac{y^2 \left(5 \left(4 y^2-3\right) \gamma +2 \left(4 y^4+3 y^2-1\right)
   m_Q \right) \alpha ^2}{4 \left(y^2+1\right)^{9/2}}\right)+\mathcal{O}\left(\varepsilon ^4\right)+\mathcal{O}\left(\alpha ^5\right)\Bigg).
\end{align}
Similarly the axion backreaction, or derivative of the axial current reads:
\begin{align}
\mathcal{B}_n = - i\frac{2 M_n}{a^3}  \frac{\lambda}{f}\langle \bar{\psi}_n\gamma^5 \psi_n \rangle = \frac{2 M_n}{a^3}  \frac{\lambda}{f}\int \frac{d^3 k}{(2\pi)^3} \sum_j \bar{V}_{j , n}\gamma^i \tau^a V_{j, n} = \frac{H^4}{\pi^2 \varepsilon^4}  \frac{\lambda}{f}\int y^2 d\ln y \, {\rm B}_n(y),
\end{align}
where  (using \textsc{Mathematica}) we find
\begin{align}
{\rm B}_n(x) =&\frac{3  \alpha  \gamma  y^2}{\left(y^2+1\right)^{5/2}} \\\nn 
&-\frac{5  \alpha  \gamma  \left(12 m_Q ^2 \left(y^3+y\right)^2+y^2 \left(\alpha ^2 \left(20
   y^4-37 y^2+6\right)-4 \gamma ^2 \left(4 y^4+y^2-3\right)\right)\right)}{8
   \left(y^2+1\right)^{11/2}}.
\end{align}
In these expressions, $\alpha = H/T$ is an order counting parameter for the WKB approximation. At the end of the computations $\alpha \to 1$. We keep this parameter in these expressions to indicate the origin (in the WKB expansion) of each of the terms in the contribution to the current. We verified that no corrections to the above survive the limit $\varepsilon \to 0$ after integration out to fifth order in the expansion in $\varepsilon$ and fifth order in the WKB approximation.

\section{Analytic integrations}\label{app:analints}

In this appendix we outline the analytic integrals over the products of Whittaker functions that appear above. The integrals are divergent, and here we regulate them with a hard cut-off. The dependence on this cutoff is cancelled by the regularization procedures discussed above in the text.

\subsection*{Axial current}
The contribution of the decoupled modes to the backreaction on the axion reads
\begin{align}\nn
\mathcal B_{\rm decoupled}  =  &  \mu \frac{\lambda}{f}\( 2 \int dx\, x^2  \Im{(u_1^{+*} v_1^+ -u_2^{-*} v_2^-) } \) \\
 & =  \mu \frac{\lambda}{f}\Im \[  i\mu e^{\pi\tilde\kappa} \int dx\, x \, W_{-\frac12+ i \tilde\kappa, i\tilde{\mu}}(2ix)W_{\frac12-i \tilde\kappa,i\tilde{\mu}}(-2ix) - (\tilde{\kappa} \to -\tilde{\kappa})\]\\
 & =   \mu \frac{\lambda}{f}\Im \[  \mathcal{I}_1 - \mathcal{I}_2 \].
\end{align}
Once regulated, the integrals over the Whittaker functions can be computed analytically.

We start by rewriting the Whittaker function in the Mellin-Barnes representation, 
\begin{align}
W_{\alpha,\beta}\left(z\right)=\frac{e^{-\frac{1}{2}z}}{2\pi{i}}\*\int_{-%
{i}\infty}^{{i}\infty}\frac{\Gamma\left(\frac{1}{2}+\beta+t\right)%
\Gamma\left(\frac{1}{2}-\beta+t\right)\Gamma\left(-\alpha-t\right)}{\Gamma\left(%
\frac{1}{2}+\beta-\alpha\right)\Gamma\left(\frac{1}{2}-\beta-\alpha\right)}z^{-t}%
\mathrm{d}t,
\end{align}
where the contour of integration is chosen to separate the poles in $\Gamma\left(\frac{1}{2}+\beta+t\right)%
\Gamma\left(\frac{1}{2}-\beta+t\right)$ from those in $\Gamma\left(-\alpha-t\right)$.

Inserting this expansion into the integral above, we get
\begin{align}\nn
\mathcal{I}_1(\Lambda)&= i\mu e^{\pi\tilde\kappa} \int dx \,x  \*\int_{%
{i}\infty}^{-{i}\infty}\frac{\Gamma\left(\frac{1}{2}-i \tilde\mu+t\right)%
\Gamma\left(\frac{1}{2}+i\tilde\mu+t\right)\Gamma\left(-1/2+i\tilde\kappa -t\right)}{\Gamma\left(%
-i\tilde\mu+i\tilde\kappa\right)\Gamma\left(+i\tilde\mu+i\tilde\kappa\right)}(-2ix)^{-t}%
\mathrm{d}t\\&~ \times\*\int_{-{i}\infty}^{{i}\infty}\frac{\Gamma\left(\frac{1}{2}+i \tilde\mu+s\right)%
\Gamma\left(\frac{1}{2}-i\tilde\mu+s\right)\Gamma\left(1/2-i\tilde\kappa -s\right)}{\Gamma\left(%
1+ i\tilde\mu-i\tilde\kappa\right)\Gamma\left(1-i\tilde\mu-i\tilde\kappa\right)}(2ix)^{-s}%
\mathrm{d}s.
\end{align}
We regulate the integral with  hard UV and IR cutoffs,  exchange the order of integrations and integrate over $x$
\begin{align}\nn
\mathcal{I}_1(\Lambda) &=- \frac{i\mu e^{\pi\tilde\kappa}}{(2\pi)^2} \int_{-i\infty}^{i\infty} ds \int_{-i\infty}^{i\infty}dt\, \frac{\Gamma\left(\frac{1}{2}-i \tilde\mu+t\right)\Gamma\left(\frac{1}{2}+i\tilde\mu+t\right)\Gamma\left(-1/2+i\tilde\kappa -t\right)}{\Gamma\left(%
-i\tilde\mu+i\tilde\kappa\right)\Gamma\left(+i\tilde\mu+i\tilde\kappa\right)}\\
&\quad ~\times\frac{\Lambda^{2-t-s}}{2-t-s} 2^{-t-s} i^{t-s} \*\frac{\Gamma\left(\frac{1}{2}+i \tilde\mu+s\right)%
\Gamma\left(\frac{1}{2}-i\tilde\mu+s\right)\Gamma\left(1/2-i\tilde\kappa -s\right)}{\Gamma\left(%
1+ i\tilde\mu-i\tilde\kappa\right)\Gamma\left(1-i\tilde\mu-i\tilde\kappa\right)}.
\end{align}
The remaining integrations can then be performed using the residues theorem (for the details of a similar computation, see the appendix of Ref.\ \cite{Adshead:2018oaa}), to find
\begin{align}
\Im \[ \mathcal{I}_1(\Lambda) \]
& =  \frac{1}{2} \mu\Lambda + \frac{3 \tilde\kappa\mu }{2} \log (2\Lambda )  + \mu\left(-\frac{\pi  \mu ^2}{4}+\frac\pi2  \tilde\kappa ^2+\frac32 \gamma  \tilde\kappa -\frac{15 \tilde\kappa }{4}-\frac{\pi }{4}\right) \\\nn
&\quad+\mu \sum_{b=\pm} \bigg\{\frac{1}{2}\Im\left[\left(\mu ^2-2 \tilde\kappa ^2-3 i \tilde\kappa +1\right) H_{i \left(- \tilde\kappa+b\sqrt{\mu ^2+ \tilde\kappa ^2} \right)}\right] \\\nn
&\qquad+\left[e^{\pi\left(   \tilde\kappa - b \sqrt{\mu ^2+ \tilde\kappa ^2}\right)} \sinh \left(\pi  \left( \tilde\kappa+b\sqrt{\mu ^2+\tilde\kappa ^2} \right)\right) \text{csch}\left(2 \pi b \sqrt{\mu ^2+ \tilde\kappa ^2}\right)\right]\\\nn
&\qquad\times\left(\frac{3b}{4} \sqrt{\mu ^2+ \tilde\kappa ^2}+2 \tilde\kappa-\frac{1}{2}\Im\left[\left(\mu ^2-2 \tilde\kappa ^2-3 i \tilde\kappa +1\right) H_{i \left(- \tilde\kappa+b\sqrt{\mu ^2+ \tilde\kappa ^2} \right)}\right]  \right) \bigg\}.
\end{align}
We can then find the total contribution from the decoupled sectors to the (regulated) axial current
\begin{align}
\mathcal B_{\rm decoupled}(\Lambda) &= 2\mu^2 \frac{\lambda}{f} \frac{H^4}{2\pi^2} \bigg\{  \left[{3 \tilde\kappa } \log (2\Lambda )  + \tilde\kappa\left(3\gamma  -{15/2  } \right)\right]\\\nn
&\quad+ \sum_{r, b=\pm} \bigg\{\frac{1}{2}\Im\left[\left(\mu ^2-2 \tilde\kappa ^2-3 i r \tilde\kappa +1\right) H_{i \left(- r\tilde\kappa+b\sqrt{\mu ^2+ \tilde\kappa ^2} \right)}\right] \\\nn
&\quad+\left[e^{\pi\left(  r \tilde\kappa - b \sqrt{\mu ^2+ \tilde\kappa ^2}\right)} \sinh \left(\pi  \left( r\tilde\kappa+b\sqrt{\mu ^2+ \tilde\kappa ^2} \right)\right) \text{csch}\left(2 \pi b \sqrt{\mu ^2+\tilde\kappa ^2}\right)\right]\\\nn
&\quad\times\left(\frac{3b}{4} \sqrt{\mu ^2+ \tilde\kappa ^2}+2 r\tilde\kappa-\frac{1}{2}\Im\left[\left(\mu ^2-2 \tilde\kappa ^2-3 i r\tilde\kappa +1\right) H_{i \left(-r \tilde\kappa+b\sqrt{\mu ^2+ \tilde\kappa ^2} \right)}\right]  \right) \bigg\} \bigg\}.
\end{align}

\subsection*{Gauge current}

Making use of the same technique, we can also compute the contribution of the decoupled modes to the gauge current. This reads
\begin{align}
\mathcal J_{\rm decoupled} =\frac{H^3}{6} \frac{1}{2\pi^2}   \int dx\, x^2  \[  ( |{u_1^+}|^2 - |{v_1^+}|^2) + ( |{u_2^-}|^2 - |{v_2^-}|^2)\]. 
\end{align}
Note that the normalization of the mode solutions allow us to write this
\begin{align}
\mathcal J_{\rm decoupled} =-\frac{H^3}{6} \frac{1}{2\pi^2} \int dx\, x^2  \[  |{v_1^+}|^2 -  |{u_2^-}|^2 \]. 
\end{align}
Inserting the Mellin-Barnes expansion, regulating with a hard cutoff, we obtain
\begin{align}\nn
 \mathcal J_{\rm decoupled}(\Lambda) 
  & = \frac{\delta_i^a}{ \eta^3 a^5} \sqrt{\frac{2}{\pi}} \Bigg[4\mu^2   \tilde\kappa \log (2\Lambda )+4 \gamma  \mu ^2  \tilde\kappa -7 \mu ^2  \tilde\kappa +\frac{8 \tilde\kappa^3}{3}-\frac{ 4 \tilde\kappa }{3} \\\nn
  &+2 \mu ^2  \tilde\kappa  \sum_{r=\pm1}\sum_{b=\pm1} \bigg\{\Re\[H_{i (r  \tilde\kappa + b  \sqrt{\mu ^2+  \tilde\kappa ^2})-1}\]+\text{csch}\left(2 \pi  \sqrt{\mu ^2+ \tilde\kappa ^2}\right)\\\nn
&\qquad  \times \sinh \left(  -\pi\sqrt{\mu ^2+ \tilde\kappa^2}-r\pi  \tilde\kappa  \right) e^{\pi  b \left(\sqrt{\mu ^2+ \tilde\kappa ^2}-r \tilde\kappa  \right)} \Re\left[H_{i b \left( r  \tilde\kappa - \sqrt{\mu ^2+ \tilde\kappa ^2}\right)+2}\right] \bigg\}\\
  &-2 \text{csch}\left(2 \pi  \sqrt{\mu ^2+  \tilde\kappa ^2}\right) \left({C_-} \sinh (2 \pi \tilde\kappa )-{C_+} \sinh \left(2 \pi  \sqrt{\mu ^2+ \tilde\kappa ^2}\right)\right)\Bigg],
\end{align}
where
\begin{align}
&C_-=\left(\frac{1}{3} \left(2 \tilde{\kappa} ^2+\frac{\mu ^2}{2}-1\right)-\frac{\left(\mu ^2+1\right)^2 \mu ^2}{4 \tilde{\kappa} ^2+\left(\mu ^2+1\right)^2}-\frac{ \left(\mu ^2+4\right)^2\mu ^2}{2 \left(16 \tilde{\kappa} ^2+\left(\mu ^2+4\right)^2\right)}\right) \sqrt{\tilde{\kappa} ^2+\mu ^2},\\
&C_+=  \left( \frac{(\mu ^4-16)\mu ^2}{2(16 \tilde{\kappa} ^2+\left(\mu ^2+4\right)^2)} +\frac{\left(\mu ^4-1\right)\mu ^2}{4 \tilde{\kappa} ^2+\left(\mu ^2+1\right)^2}-\frac{2 \tilde{\kappa} ^2}{3}+\frac{1}{3}\right)\tilde{\kappa}.
\end{align}

\paragraph{Massless limit gauge current}

In the massless limit, we can compute the contribution of the coupled modes to the gauge current
\begin{align}
\mathcal J_{\mu = 0} = \frac{H^3 }{ 6} {\frac{1}{2\pi^2}} \int dx\, x^2  \[  ( |{u_1^+}|^2 - |{v_1^+}|^2) + ( |{u_2^-}|^2 - |{v_2^-}|^2) + 8 \Re{(v_1^{- *} v_2^{+}  )}  \]. 
\end{align}
The integral for the decoupled sector vanishes when $\mu = 0$. The remaining integral is
\begin{align}
& \int dx\, x^2  \[ 8 \Re{(v_1^{- *} v_2^{+}  )}  \]
=- 8 m_Q \int dx\,  \Im{\[ x\, W_{-\frac12, im_Q}(2ix) W_{\frac12, im_Q}(-2ix)\]},
\end{align}
which, after regulation with a hard cutoff, evaluates to
\begin{align}\nn
\int dx\, x^2  \[ 8 \Re{(v_1^{- *} v_2^{+}  )}  \]&=4 \left( m_Q\Lambda ^2 - \left(m_Q^3+m_Q\right)\log (2 \Lambda )\right)\\
& +2 \left(2 \left(m_Q^3+m_Q\right) \left(-\gamma +\Re\left[H_{-i \sqrt{m_Q^2}}\right]\right)+m_Q^3+3m_Q\right).
\end{align}
So, we find the full gauge current in the massless limit is
\begin{align}\nn
 \mathcal J_{\rm decoupled}(\Lambda) 
  & = \frac{H^3 }{ 6} {\frac{1}{2\pi^2}} \Bigg[4 \left( m_Q\Lambda ^2 - \left(m_Q^3+m_Q\right)\log (2 \Lambda )\right)\\
& +2 \left(2 \left(m_Q^3+m_Q\right) \left(-\gamma +\Re\left[H_{-i \sqrt{m_Q^2}}\right]\right)+m_Q^3+3m_Q\right)
\Bigg].
\end{align}
Note that this is independent of the axion.

\bibliographystyle{JHEP}
\bibliography{references}
\end{document}